\documentclass[showpacs,twocolumn,aps,floatfix,superscriptaddress,noshowpacs]{revtex4}

\usepackage{color}
\usepackage{psfrag}
\usepackage{ulem,cancel}

\usepackage{amsthm,braket,here}
\usepackage{amsmath}
\usepackage{amsfonts}
\usepackage{amssymb}
\usepackage{enumitem}
\usepackage{comment}

\usepackage[utf8]{inputenc}
\usepackage{t1enc}
\usepackage[mathlines]{lineno}% Enable numbering of text and display math

\usepackage{amsthm,amsmath,amssymb,graphicx,bm,color,mathrsfs,verbatim,epstopdf,dcolumn,dsfont}
\usepackage{comment}
\graphicspath{{figs/},{./}}

\usepackage{hyperref}
\hypersetup{
	colorlinks   = true
}

\newcommand{\be}{\begin{equation}}
\newcommand{\ee}{\end{equation}}

\newcommand{\av}[1]{{\overline{#1}}}
\newcommand{\mc}[1]{\mathcal{#1}}
\newcommand{\mr}[1]{\mathrm{#1}}

\newcommand{\EQ}[1]{Eq.~\eqref{#1}}

\newcommand{\hH}{\hat{H}}

\newcommand{\NON}{\nonumber\\}
\newcommand{\hrho}{\hat{\rho}}
\newcommand{\lrs}[1]{\left( #1 \right)}

\newcommand{\lrl}[1]{\left[ #1 \right]}

\newcommand{\aln}[1]{
\begin{align}
#1
\end{align}
}

\newcommand{\ra}{\rightarrow}

\newcommand{\Tr}{\mr{Tr}}
\newcommand{\hmo}{\hat{\mc{O}}}

\begin{document}

%\preprint{APS/123-QED}

%\title{Nonequlibrium stationary states driven by the Maxwell's Demon in an optical lattice}
\title{Random-matrix behavior of quantum nonintegrable many-body systems with Dyson's three symmetries}
%\\

\author{Ryusuke Hamazaki}
\email{hamazaki@cat.phys.s.u-tokyo.ac.jp}
\affiliation{Department of Physics, University of Tokyo, 7-3-1 Hongo, Bunkyo-ku, Tokyo 113-0033, Japan}

\author{Masahito Ueda}
\affiliation{Department of Physics, University of Tokyo, 7-3-1 Hongo, Bunkyo-ku, Tokyo 113-0033, Japan}
\affiliation{RIKEN Center for Emergent Matter Science (CEMS), Wako 351-0198, Japan}

\begin{abstract}
We propose a one-dimensional nonintegrable spin model with local interactions that covers Dyson's three symmetry classes (classes A, AI, and AII) depending on the values of parameters.
We show that the nearest-neighbor spacing distribution in each of these classes  agrees with that of random matrices with the corresponding symmetry.
By investigating the ratios between the standard deviations of diagonal and off-diagonal matrix elements,
we numerically find that they become universal, depending only on symmetries of the Hamiltonian and an observable, as predicted by random matrix theory.
These universal ratios are evaluated from long-time dynamics of small isolated quantum systems.

\end{abstract}

\date{\today}
%\pacs{67.85.Pq, 71.10.Fd}
\maketitle

\section{\label{sec:intro}Introduction}
Studies on quantum many-body systems over decades have uncovered universal behaviors in such systems upon breaking of integrability.
Originally investigated in the context of  nuclei~\cite{Wigner51,Brody81,Haq82,Zelevinsky96,Weidenmuller09,Mitchell10}, atoms~\cite{Rosenzweig60,Camarda83} and molecules~\cite{Haller83,Abramson84}, such quantum nonintegrable many-body systems are conjectured to have the eigenenergy and eigenstate statistics described by random matrices~\cite{semi-Master}.
Recently, experiments of ultracold atoms and trapped ions have motivated theorists to study random-matrix behaviors of locally interacting systems of bosons, fermions, and spins from   numerical~\cite{Karthik07,Santos10a,Pal10,Khatami13,Beugeling14,Kim14,Hamazaki16G,Beugeling15,DAlessio16,Bera16,Luitz16,Serbyn16S,Mondaini16,Mondaini17,Pal18,Khaymovich18,Hamazaki18N} and analytical~\cite{Erdos14,Keating14,Cunden17,Kos17,Chan18,Bertini18} perspectives.
For example, matrix elements of an observable with respect to energy eigenstates of a nonintegrable Hamiltonian are demonstrated to be well described by random matrices.
 This fact is closely related to the eigenstate thermalization hypothesis (ETH)~\cite{Neumann29,Jensen85,Deutsch91,Srednicki94,Tasaki98,Rigol08}, which states that each energy eigenstate behaves thermal for typical observables.
The ETH in many-body systems is numerically verified in various nonintegrable systems and considered to lay the cornerstone for universal aspects of statistical mechanics, such as thermalization~\cite{Rigol08,Santos10a,Santos10b,Ikeda11,Khodja15,DAlessio16,Kaufman16,Clos16,Reimann16,Mori18}, information scrambling~\cite{Huang17} and irreversibility~\cite{Schmitt17,Hamazaki18O}.

Universal properties of quantum many-body systems are classified by antiunitary symmetry that commutes with the Hamiltonian, in particular time-reversal symmetry.
Dyson~\cite{Dyson62T} introduced the three fundamental symmetry classes: (i) class A which does not have antiunitary symmetry, (ii) class AI which has antiunitary symmetry $\hat{T}$ that satisfies $\hat{T}^2=1$ and $[\hat{T},\hat{H}]=0$,
and (iii) class AII which has antiunitary symmetry $\hat{T}$ that satisfies $\hat{T}^2=-1$ and $[\hat{T},\hat{H}]=0$.
Gaussian random-matrix ensembles in each of these classes are called (i) the Gaussian unitary ensemble (GUE), (ii) the Gaussian orthogonal ensemble (GOE), and (iii) the Gaussian symplectic ensemble (GSE).
It is expected that eigenenergy and eigenstate statistics of quantum nonintegrable many-body systems are understood from those of
random matrices that belong to the same symmetry class.

A theoretical study of universality in nonintegrable many-body systems
requires appropriate  models,
since too complicated models are often intractable and  too simplified models may have extra symmetry and integrability.
For example, while the transverse-field Ising model in one dimension is integrable, an additional longitudinal field to this model makes it nonintegrable~\cite{Banuls11}.
Indeed, the eigenenergy and eigenstate statistics of this two-field Ising model are described by GOE because the model belongs to class AI.
This model consists of only local and up to two-body interactions, and has been used to study quantum chaos, thermalization, etc., numerically~\cite{Shenker14,Zhang15,Kim15,Hosur16} and experimentally~\cite{Smith16}.
On the other hand, currently known nonintegrable many-body models with local interactions do not  cover all of the three symmetry classes, especially class AII.
For studying the effects of symmetry, it is desirable to have models whose symmetry can be controlled by parameters of the Hamiltonian.

In this paper~\cite{added-Master}, we propose a one-dimensional nonintegrable spin model that covers Dyson's three symmetries (classes A, AI, and AII) depending on the values of parameters.
This model is composed of up to two-body local interactions including the Dzyaloshinskii-Moriya (DM) interaction~\cite{Dzyaloshinsky58,Moriya60} to describe class AII.
We study nearest-neighbor spacing distributions of models in these classes and show that they obey those of random matrices with the corresponding symmetry (GUE, GOE, and GSE).
At intermediate parameters, crossover transitions of the distributions are observed.
We also study the ratios between the standard deviations of diagonal and off-diagonal matrix elements.
We show that the ratios become universal, depending only on symmetries of the Hamiltonian and an observable, as predicted by random matrix theory.
These universal ratios are evaluated from long-time dynamics of small isolated quantum systems.

The rest of this paper is organized as follows.
In Sec.~\ref{sec:model}, we define three nonintegrable models with different symmetries.
In Sec.~\ref{sec:level}, we analyze the nearest-neighbor spacing distributions of eigenenergies for  these models.
We also demonstrate that crossover transitions between these models occur at intermediate parameters.
In Sec.~\ref{sec:ratio}, we investigate the ratios between the standard deviations of diagonal and off-diagonal matrix elements.
After introducing the predictions on these ratios using random matrix theory, we numerically show that the predictions hold true for our models and observables with various symmetries.
In Sec.~\ref{sec:quench}, we analyze two types of quench dynamics in our models to demonstrate that the universal ratios are evaluated from long-time dynamics of autocorrelation functions and temporal fluctuations.
In Sec.~\ref{sec:outro}, we summarize the main results of this work and discuss some outlooks.
In appendices, we study another indicator for the level-spacing statistics and prove several relations used in the main text.

\section{\label{sec:model}Model and its symmetries}
We introduce a one-dimensional spin model with local interactions, which changes its symmetry by varying parameters.
Our model consists of the Ising interaction, transverse and longitudinal fields and the DM interaction as follows:
\aln{\label{spinchain}
\hat{H}&=\hat{H}_\mr{I}+\hat{H}_\mr{F}+\hat{H}_\mr{DM},\\
\hat{H}_\mr{I}&= -\sum_{i=1}^{N-1}J(1+\epsilon_{i})\hat{\sigma}_i^z\hat{\sigma}_{i+1}^z,\\
\hat{H}_\mr{F}&= -\sum_{i=1}^{N}(h'\hat{\sigma}_i^x+h\hat{\sigma}_i^z),\\
\hat{H}_\mr{DM}&=\sum_{i=1}^{N-1}\vec{D}\cdot (\vec{\hat{\sigma}}_i\times\vec{\hat{\sigma}}_{i+1})\NON
&=\frac{D}{\sqrt{2}}\sum_{i=1}^{N-1}[\lrs{\hat{\sigma}_i^y\hat{\sigma}_{i+1}^z-\hat{\sigma}_i^z\hat{\sigma}_{i+1}^y}\NON
&\:\:\:\:\:\:\:\:\:+
\lrs{\hat{\sigma}_i^x\hat{\sigma}_{i+1}^y-\hat{\sigma}_i^y\hat{\sigma}_{i+1}^x}
],
}
where $N$ denotes the number of spins, $\vec{D}=D\frac{1}{\sqrt{2}}(\vec{e}_x+\vec{e}_z)$, and we impose the open boundary condition.
In addition, $\epsilon_i$ is a random variable that is uniformly chosen from $[-\epsilon,\epsilon]$ at each site to break the reflection symmetry of sites $i\leftrightarrow N-i$.
As we will see in the discussions below, the randomness is sufficiently weak and no localization arises.

Let us explain each term in \EQ{spinchain}.
The Ising Hamiltonian $\hH_\mr{I}$ has anisotropy in spin space, which makes it less symmetric compared with isotropic interactions, e.g., Heisenberg interactions.
By adding transverse and longitudinal fields $\hH_\mr{F}$, the Hamiltonian $\hH_\mr{I}+\hH_\mr{F}$ loses unitary symmetry that commutes with it.
On the other hand, it has one antiunitary symmetry
$\hat{T}=\hat{K}\:(\hat{T}^2=1)$, where $\hat{K}$ denotes complex conjugation and satisfies
\aln{
\hat{K}\hat{\sigma}_i^x\hat{K}^{-1}=\hat{\sigma}_i^x, \hat{K}\hat{\sigma}_i^y\hat{K}^{-1}=-\hat{\sigma}_i^y, \hat{K}\hat{\sigma}_i^z\hat{K}^{-1}=\hat{\sigma}_i^z.
}

Next, to realize systems with antiunitary symmetry satisfying $\hat{T}^2=-1$,
we note that $\hH_\mr{I}$ has time-reversal symmetry
\aln{
\hat{T}=\hat{T}_0:=\lrs{\prod_{i=1}^{N}[i\hat{\sigma}_i^y]}\hat{K},
}
which satisfies $\hat{T}_0^2=(-1)^N$ and
\aln{
\hat{T}_0\hat{\sigma}_i^x\hat{T}_0^{-1}=-\hat{\sigma}_i^x, \hat{T}_0\hat{\sigma}_i^y\hat{T}_0^{-1}=-\hat{\sigma}_i^y, \hat{T}_0\hat{\sigma}_i^z\hat{T}_0^{-1}=-\hat{\sigma}_i^z.
}
To keep this symmetry, we are only allowed to add even-body interactions of spins to $\hat{H}_\mr{I}$.
To break other symmetries such as $\hat{K}$, we choose the DM (DM) interaction $\hH_\mr{DM}$.
Indeed, $\hH_\mr{I}+\hH_\mr{DM}$ with $D\neq 0$ only has $\hat{T}_0$ as symmetry.
Note that interactions of the form $\hat{\sigma}_i^\alpha\hat{\sigma}_{i+1}^\alpha\:(\alpha=x,y,z)$ do not work because they cannot break symmetries such as $\hat{K}$.
The DM interaction with $\vec{D}\propto \vec{e}_\alpha\:(\alpha=x,y,z)$ does not work for the same reason, either.

Since $\hH_\mr{F}$ and $\hH_\mr{DM}$ have different antiunitary symmetry, $\hH_\mr{I}+\hH_\mr{F}+\hH_\mr{DM}$ for nonzero $h,h'$ and $D$ is not constrained by any symmetry.
Thus, by varying $h,h'$ and $D$ we obtain models belonging to different symmetry classes.
We particularly introduce the following three models, whose parameters are chosen such that
their unwanted symmetries and integrability are broken:
\aln{\label{modela}
\mr{model\:a}:\: \hH_\mr{a}=\hH_\mr{I}+\hH_\mr{F}+\hH_\mr{DM}
}
with $h=0.5, h'=-1.05$ and $D=0.9$,
\aln{\label{modelb}
\mr{model\:b}:\: \hH_\mr{b}=\hH_\mr{I}+\hH_\mr{F}
}
with $h=0.5, h'=-1.05$ and $D=0$, and
\aln{\label{modelc}
\mr{model\:c}:\: \hH_\mr{c}=\hH_\mr{I}+\hH_\mr{DM}
}
with $h=h'=0$ and $D=0.9$.
We also assume that $J=1$ and $\epsilon=0.1$ in $\hH_\mr{I}$ in all of the models.
Model a belongs to class A and model b belongs to class AI.
On the other hand, model c belongs to class AI for even $N$ and class AII for odd $N$, since $\hat{T}_0^2=(-1)^N$.
%In Sec.~\ref{sec:level}, we also discuss the property of models with intermediate parameters between these three models.

\section{\label{sec:level}Nearest-neighbor level-spacing distributions}
\begin{figure}
\begin{center}
\includegraphics[width=\linewidth]{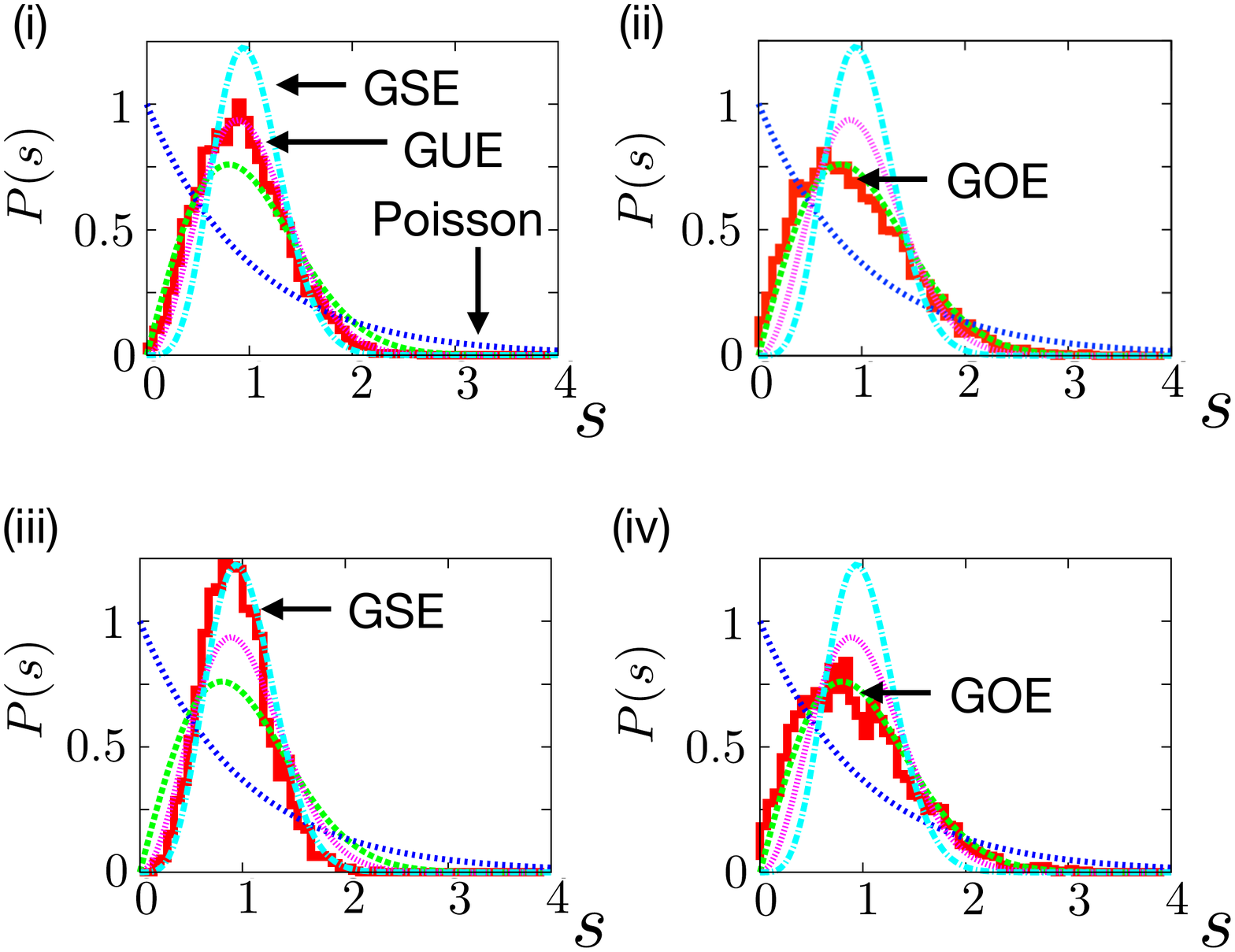}
\caption{
Level-spacing distributions $P(s)$ for (i) model a with $N=13$, (ii) model b with $N=13$, (iii) model c with $N=13$, and (iv) model c with $N=12$ for the Hamiltonians in Eqs.~\eqref{modela}-\eqref{modelc}.
Superimposed are the level-spacing statistics for the Poisson ($P_\mr{P}(s)=e^{-s}$), GUE~\eqref{psgue}, GOE~\eqref{psgoe}, and GSE~\eqref{psgse}.
We clearly see that the data in (i), (ii), (iii) and (iv) are consistent with GUE, GOE, GSE and GOE, respectively.
We use 20\% of the total number of the unfolded eigenenergies from the middle of the spectrum (by identifying the Kramers pairs), i.e., $1.6\times 10^3$ eigenenergies for (i) and (ii), and $8\times 10^2$ eigenenergies for (iii) and (iv).
}
\label{fig1}
\end{center}
\end{figure}

We first analyze nearest-neighbor level-spacing distributions $P(s)$~\cite{Haake}, which are primary indicators of universality in nonintegrable systems.
Figure~\ref{fig1}(i) shows that the distribution for model a defined in the previous section obeys GUE, \aln{\label{psgue}
P_\mr{GUE}(s)=\frac{32}{\pi^2}s^2e^{-\frac{4}{\pi}s^2},
}
 reflecting the fact that this model belongs to class A.
 Figure~\ref{fig1}(ii) shows that the distribution for model b obeys GOE,
\aln{\label{psgoe}
P_\mr{GOE}(s)=\frac{\pi }{2}se^{-\frac{\pi }{4}s^2},
}
consistent with the fact that this model belongs to class AI.
Finally, the distribution for model c varies according to the parity of $N$ as shown in Fig.~\ref{fig1}(iii) and (iv): it obeys GSE,
\aln{\label{psgse}
P_\mr{GSE}(s)=\frac{2^{18}}{3^6\pi^3}s^4e^{-\frac{64}{9\pi}s^2}
}
 for odd $N$ and GOE for even $N$.
 Note that the nearest-neighbor level-spacing distribution for odd $N$ (class AII) in Fig.~\ref{fig1} is calculated by identifying the two degenerate eigenenergies, i.e., the Kramers pairs.
These results indicate that our models are in fact sufficiently nonintegrable and that their eigenenergy statistics are described by random matrix theory that respects the corresponding symmetry.
We also calculate another indicator of level spacing distributions, i.e., the ratio for consecutive level spacings~\cite{Atas13} and confirm that it also agrees with the prediction of random matrix theory, as shown in Appendix~\ref{app1}.

\begin{figure*}
\begin{center}
\includegraphics[width=\linewidth]{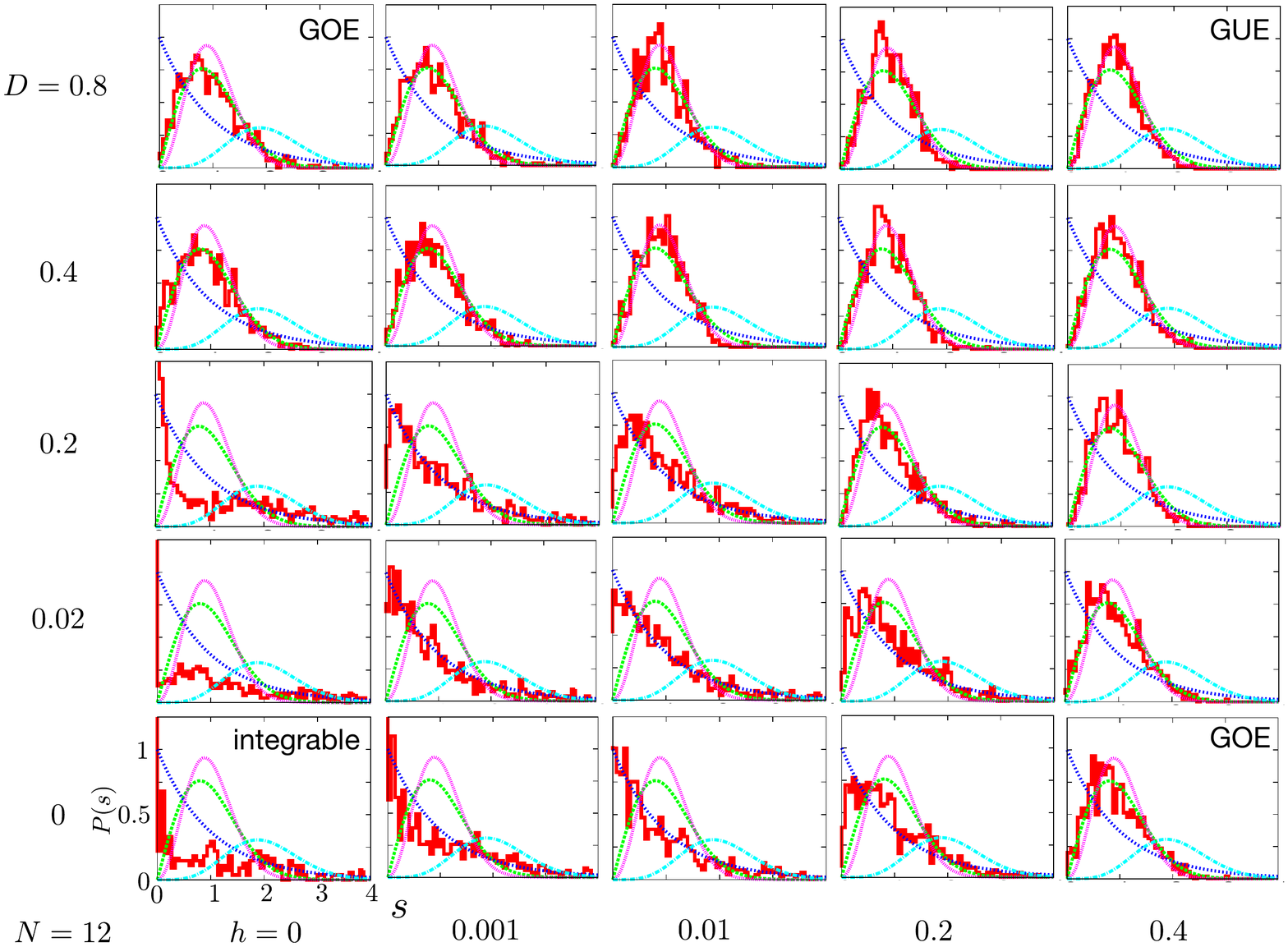}
\caption{
Nearest-neighbor level-spacing distributions $P(s)$ for Hamiltonians~\eqref{spinchain}  with $N=12$, $J=1$ and $h'=-2.1h$.
For small $h$ and $D$, $P(s)$ does not show level repulsions because the system is still approximately integrable.
For  $h=0$ and large $D$ or for $D=0$ and large $h$, $P(s)$ is close to GOE statistics,
and for large $h$ and $D$,  $P(s)$ is close to GUE.
We find crossover transitions between these different statistics.
We use 20\% of the total number of the unfolded eigenenergies from the middle of the spectrum, i.e., about $8\times 10^2$ eigenenergies.
}
\label{fig6}
\end{center}
\end{figure*}

Moreover, we find that the nearest-neighbor spacing distributions exhibit crossover transitions among different symmetries if we change the parameters of our Hamiltonian in Eq.~\eqref{spinchain}.
Figure~\ref{fig6} shows the nearest-neighbor level-spacing distributions $P(s)$ for Hamiltonians~\eqref{spinchain}  with $N=12$ by varying parameters.
We fix $J=1$ and $h'=-2.1h$ and vary $h$ and $D$.
For small $h$ and $D$, the system is still approximately integrable and $P(s)$ does not show level repulsions.
Note that the exact degeneracies due to the global spin flip for $h=D=0$ (purely Ising case) make $P(s)$ differ from the typical exponential form of the Poisson distribution and add the peak at $s=0$~\cite{degen-Master}.
For  $h=0$ and large $D$, or for $D=0$ and large $h$, $P(s)$ is close to GOE statistics, since $N$ is even.
For large $h$ and $D$,  $P(s)$ is close to GUE statistics because all symmetries are broken.
Between these different statistics the crossover transitions are observed.
In general, transitions are sharper for nonintegrable-nonintegrable ones (such as GOE-GUE transitions) than integrable-nonintegrable ones.

\begin{figure*}
\begin{center}
\includegraphics[width=\linewidth]{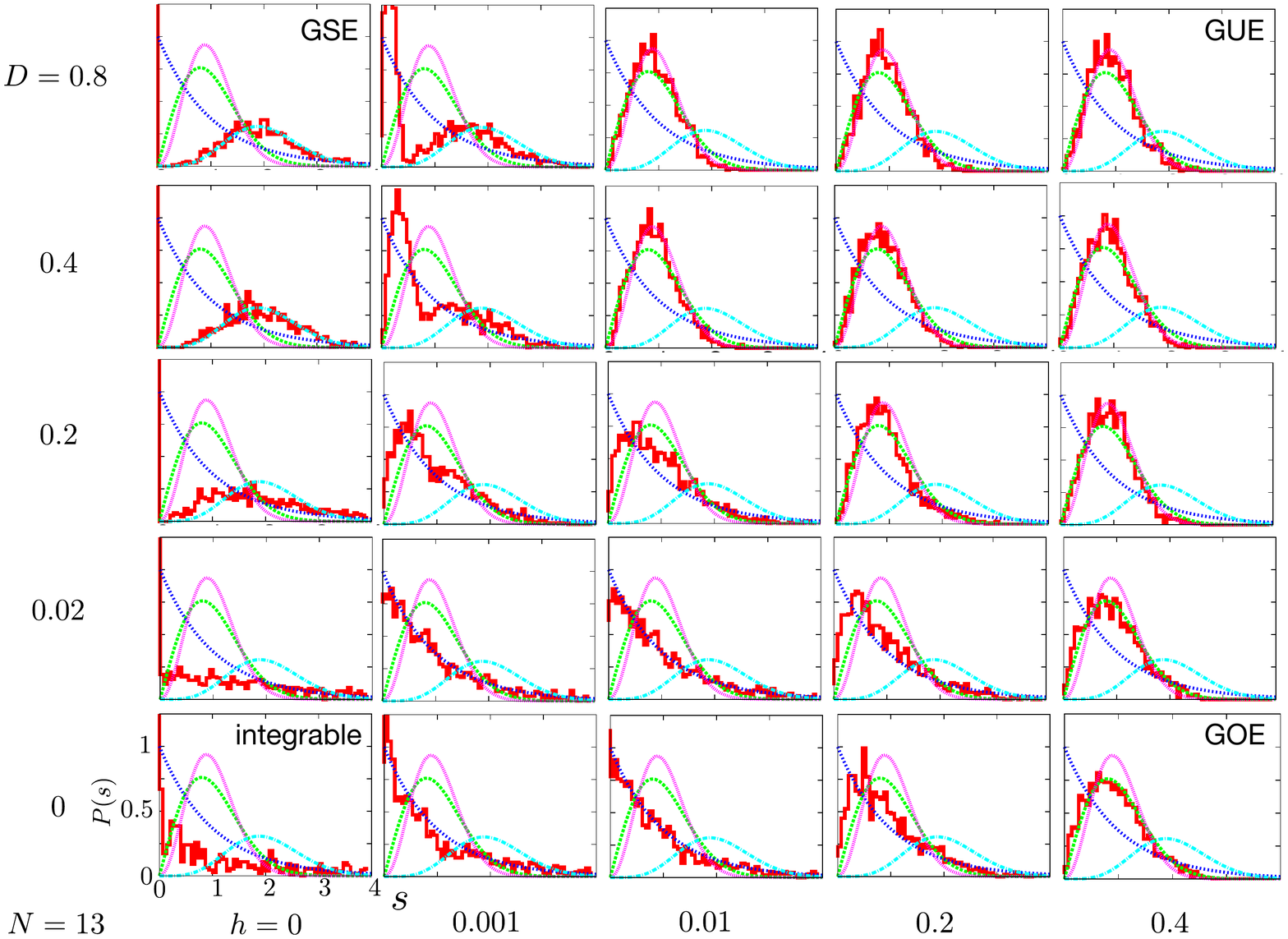}
\caption{
Nearest-neighbor spacing distributions $P(s)$ for Hamiltonians~\eqref{spinchain}  with $N=13$, $J=1$ and $h'=-2.1h$.
For small $h$ and $D$, $P(s)$ does not show level repulsions because the system is still approximately integrable.
For  $h=0$ and large $D$, $P(s)$ is close to GSE statistics.
For $D=0$ and large $h$, $P(s)$ is close to GOE statistics.
For large $h$ and $D$,  $P(s)$ is close to GUE statistics.
We find crossover transitions between these different statistics.
Note that the distribution for GSE here is written as $P(s)=\delta(s)/2+P_\mr{GSE}(s/2)/4$ using $P_\mr{GSE}(s)$ in Eq.~\eqref{psgse}, since we do not identify the Kramers pairs.
We use 20\% of the total number of the unfolded eigenenergies from the middle of the spectrum, i.e., about $1.6\times 10^3$ eigenenergies.
}
\label{fig7}
\end{center}
\end{figure*}

Similarly, Fig.~\ref{fig7} shows  $P(s)$ for Hamiltonian~\eqref{spinchain}  with $N=13$ by varying parameters with $J=1$ and $h'=-2.1h$.
For small $h$ and $D$, the system is still approximately integrable and $P(s)$ does not show level repulsions.
Note that the exact degeneracies for $h=D=0$ described above again make $P(s)$ differ from the exponential form.
For  $h=0$ and large $D$, $P(s)$ is close to GSE statistics, and for $D=0$ and large $h$, $P(s)$ is close to GOE statistics, since $N$ is odd.
For large $h$ and $D$,  $P(s)$ is close to GUE statistics because all symmetries are broken.
We again find crossover transitions between these different statistics, where transitions are sharper for nonintegrable-nonintegrable ones (such as GOE-GUE and GSE-GUE transitions) than integrable-nonintegrable ones.

\section{\label{sec:ratio}Universal ratio on matrix elements}

\subsection{\label{subsec:ratio}Ratio of the standard deviations of diagonal and off-diagonal matrix elements}

Another important universality of nonintegrable systems appears in matrix elements of observables $\hmo$ for energy eigenstates $\ket{E_\alpha}$ with eigenenergy $E_\alpha$ ($\hH\ket{E_\alpha}=E_\alpha\ket{E_\alpha}$).
To see this, we consider fluctuations of matrix elements within an energy shell $\mc{H}_{E_0,\omega_s}$, which is spanned by energy eigenstates whose eigenenergies are within $[E_0-\omega_s,E_0+\omega_s]$ for small $\omega_s$.
We especially focus on  the ratio $r$ between the standard deviation of diagonal matrix elements $\Delta\mc{O}_\mr{d}$ and that of
off-diagonal matrix elements $\Delta\mc{O}_\mr{od}$~\cite{Mondaini17,Dymarsky19}:
\aln{
r=\frac{\Delta\mc{O}_\mr{d}}{\Delta\mc{O}_\mr{od}}.
}
We can also define
\aln{
r'=\frac{\Delta\mc{O}_\mr{K}}{\Delta\mc{O}_\mr{od}}
}
for use to analyze class AII, where $\Delta\mc{O}_\mr{K}$ is the standard deviation of matrix elements within the Kramers degenerate pair $\mc{O}_{\alpha\tilde{\alpha}}=\braket{E_\alpha|\hmo|\tilde{E_\alpha}}$ with $\hH\ket{\tilde{E_\alpha}}=E_\alpha\ket{\tilde{E_\alpha}}$ and $\ket{\tilde{E_\alpha}}=\hat{T}\ket{E_\alpha}$.
Note that matrix elements $\mc{O}_{\alpha\tilde{\alpha}}$ are not invariant under rotation of two eigenstates in the two-dimensional Kramers-pair space. In other words, we can define a linear combination of $\ket{E_\alpha}$ and $\ket{\tilde{E_\alpha}}$ as a new basis. We assume that the angle of the rotation (or the coefficients of the linear combination) is uniformly randomized.

In the rest of this section, we use random matrix theory to show that these ratios become universal constants in nonintegrable systems, which depend only on the symmetries of the Hamiltonian and observables.

\subsection{\label{subsec:rmt}Random matrix theory and symmetry}
We first determine $r$ from random matrix theory.
We start from a conjecture that matrix elements $\braket{E_\alpha|\hmo|E_\beta}=\mc{O}_{\alpha\beta}$ can be written in the following form~\cite{Srednicki99,Khatami13} in nonintegrable systems with no degeneracy (i.e., for classes A and AI):
\aln{\label{sred}
\mc{O}_{\alpha\beta}=\mc{A}(E)\delta_{\alpha\beta}+\Omega(E)^{-\frac{1}{2}}f(E,\omega)R_{\alpha\beta},
}
where $E=(E_\alpha+E_\beta)/2, \omega=E_\alpha-E_\beta$, $\Omega(E)$ is the density of states of the Hamiltonian, and $\mc{A}(E)$ and $f(E,\omega)$ are smooth functions of their arguments.
In addition, $R_{\alpha\beta}$ behaves quasi-randomly as a function of $\alpha$  and $\beta$, and its statistical properties (normalized such that the average is zero and the variance is 1) are well described by random matrix theory:  $R_{\alpha\beta}$ fluctuates as if energy eigenstates were eigenstates of random Hamiltonians.
Since $\Omega(E)$ is exponentially large with respect to the size of the system, the second term in \EQ{sred} is exponentially suppressed in the thermodynamic limit, which leads to the ETH.
Equation~\eqref{sred} has numerically been verified for few-body observables~\cite{Khatami13,Beugeling14,Beugeling15,DAlessio16,Luitz16,Serbyn16S,Mondaini17,Dymarsky19}, although recent studies indicate that it may also hold for many-body observables~\cite{Khaymovich18,Hamazaki18A}.

By assuming the ansatz in \EQ{sred}, we calculate the ratio $r$ between the standard deviation of fluctuations (the second term of \EQ{sred} on the right-hand side) for diagonal matrix elements $\Delta\mc{O}_\mr{d}$ and that for
off-diagonal matrix elements $\Delta\mc{O}_\mr{od}$.
Since $\Omega(E)$ and $f(E,\omega)$ stay almost constant when the energy shell is sufficiently small (note that $|E-E_0|<\omega_s$ and $|\omega|<2\omega_s$), the ansatz leads to
\aln{\label{approximate}
r=\frac{\Delta\mc{O}_\mr{d}}{\Delta\mc{O}_\mr{od}}\simeq \frac{\Delta R_\mr{d}}{\Delta R_\mr{od}},
}
where $\Delta R_\mr{d}$ and $\Delta R_\mr{od}$ are the standard deviations of the diagonal and off-diagonal elements of $R_{\alpha\beta}$, respectively.
Since the statistics of $R_{\alpha\beta}$ are described by random matrix theory, $r$ becomes universal irrespective of the details of Hamiltonians and observables.

\begin{table}
  \begin{center}
    \caption{Universal ratios $r$ and $r'$\label{universal}. No symmetry classification of observables exists for class A and the value of $r_\mr{A}$ is formally put to the "even" place.
    We also note that $r'$ is defined only for class AII.
    }

    \begin{tabular}{ccccc} \hline\hline
     Ratio & $r_\mr{A}$ & $r_\mr{AI}$ & $r_\mr{AII}$ & $r'_\mr{AII}$ \\ \hline
      even  $\mc{\hat{O}}$& 1 & $\sqrt{2}$ & 1 & 0 \\
      odd  $\mc{\hat{O}}$& - & 0& 1 & $\sqrt{2}$ \\ \hline\hline
    \end{tabular}
    \label{table1}
  \end{center}
\end{table}

Using random matrix theory, we find that $r$ (and similarly $r'$) is actually determined only by the symmetries of the Hamiltonian and an observable, as shown in Table~\ref{table1}.
Here, the symmetry of an observable is defined by using the time-reversal symmetry $\hat{T}$ (for classes AI and AII) as follows: the symmetry of $\hmo$ is even if $\hat{T}\hmo\hat{T}^{-1}=\hmo$ and odd if $\hat{T}\hmo\hat{T}^{-1}=-\hmo$.
Note that we do not consider observables that are neither odd nor even in this paper for simplicity.
As detailed in Appendix~\ref{app2}, Table~\ref{table1} is obtained under the assumption that the eigenstates behave as if they were eigenstates of random matrices with appropriate symmetry constraints within the energy shell.

Our finding in Table~\ref{table1} extends the previous results for classes A ($r_\mr{A}=1$) and AI even ($r_\mr{AI,even}=\sqrt{2}$~\cite{Steinigeweg13,DAlessio16,Mondaini17}) to class AII.
We also note that the symmetry of the observable is crucial for the universality, which has not been discussed in previous literature.

\begin{table}
  \begin{center}
    \caption{Prediction of random matrices on matrix elements for each model and observable. Note that $r'$ is defined only for model c with odd $N$.
    }

    \begin{tabular}{ccccccc} \hline\hline
      Ratio \:& $r_\mr{a}$\: & $r_\mr{b}$\: & $r_\mr{c}$  \:& $r_\mr{c}$ \:& $r'_\mr{c}$ \\
              & &  &  (even $N$) \:&  (odd $N$) \:&  (odd $N$) \\ \hline
Symmetry        &A & AI & AI \:&  AII \:&  AII \\ \hline
      $\mc{O}_\mr{c}, \mc{O}_l\:(\mr{even}\: l)$  & 1 & $\sqrt{2}$& $\sqrt{2}$ & $1$ & 0\\
      $\mc{O}_\mr{m}, \mc{O}_l\:(\mr{odd}\: l)$ & 1 & $\sqrt{2}$ & 0 & 1 & $\sqrt{2}$\\
 \hline\hline
    \end{tabular}
    \label{table2}
  \end{center}
\end{table}

\subsection{\label{subsec:few}Numerical results for few-body observables}
We now calculate the ratios $r$ and $r'$ for our models a, b, and c explained in Sec.~\ref{sec:model} for few-body observables with different symmetries.
We especially consider the local magnetization
\aln{
\hmo_\mr{m}:=\hat{\sigma}^z_{\left[N/2\right]+1}
}
and the neighboring correlation
\aln{
\hmo_\mr{c}=\hat{\sigma}^z_{\left[N/2\right]+1}\hat{\sigma}^z_{\left[N/2\right]+2},
}
where $[x]$ is the Gauss symbol, which gives the greatest integer that does not exceed $x$.
Since $\hmo_\mr{m}$ satisfies $\hat{K}\hat{\mc{O}}_\mr{m}\hat{K}^{-1}=\hat{\mc{O}}_\mr{m}$, it is an even operator for models a and b.
On the other hand, since $\hat{T}_0\hat{\mc{O}}_\mr{m}\hat{T}_0^{-1}=-\hat{\mc{O}}_\mr{m}$, it is odd for model c.
As for $\hat{\mc{O}}_\mr{c}$, it is an even operator for all the models because $\hat{K}\hat{\mc{O}}_\mr{c}\hat{K}^{-1}=\hat{\mc{O}}_\mr{c},\hat{T}_0\hat{\mc{O}}_\mr{c}\hat{T}_0^{-1}=\hat{\mc{O}}_\mr{c}$.
If we assume the prediction of random matrix theory in Table~\ref{table1}, the ratios for matrix elements of each observable in three models are predicted as Table~\ref{table2}.

Before showing our numerical results, we comment on an important caveat in calculating $\Delta\mc{O}_\mr{d}$ to obtain $r$.
While we need to extract the second term on the right-hand side of \EQ{sred} to probe the random matrix behavior, the naive standard deviation of diagonal matrix elements involves an unwanted effect of the first term especially for few-body observables~\cite{Hamazaki18A}.
To remove this contribution, we consider the standard deviation of modified fluctuations that are defined by the deviations from the linear regression of diagonal matrix elements within the energy shell, not from the naive average (a similar analysis was performed in, e.g., Refs.~\cite{Beugeling14,IkedaD}).

\begin{figure}
\begin{center}
\includegraphics[width=\linewidth]{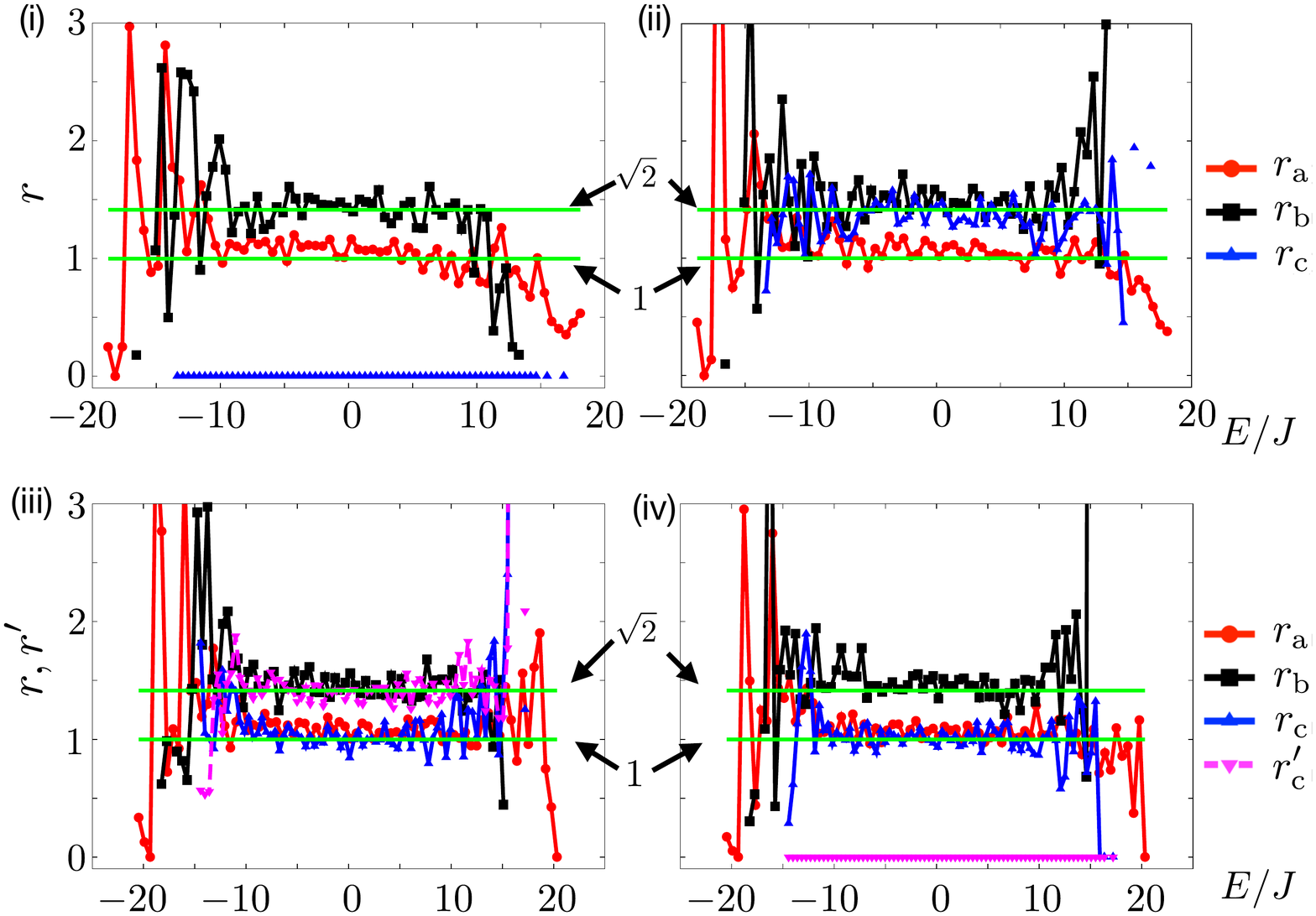}
\caption{
Ratios $r$ and $r'$ of the standard deviations of matrix elements as a function of energy.
The results are shown for (i) $\hat{\mc{O}}_\mr{m}$ with $N=12$, (ii) $\hat{\mc{O}}_\mr{c}$ with $N=12$, (iii) $\hat{\mc{O}}_\mr{m}$ with $N=13$, and (iv) $\hat{\mc{O}}_\mr{c}$ with $N=13$.
Each graph shows $r$ for models a (circle), b (square), and c (upward triangle).
For $N=13$, we also show $r'$ for model c (downward triangle).
By analyzing the data from a viewpoint of symmetry, we can see that the  conjecture in Tables~\ref{table1} and~\ref{table2} is valid for these few-body operators especially in the middle of the spectra.
We note that some data points are missing for the edges of the spectrum because only few eigenstates exist there.
Each datapoint is obtained from the standard deviation of the modified fluctuations (see the main text) within the small energy shell $\omega_s=(\max_\alpha E_\alpha-\min_\alpha E_\alpha)/(12N)$ for the Hamiltonian with a single disorder realization.
}
\label{fig2}
\end{center}
\end{figure}

In Fig.~\ref{fig2}, we show the energy dependences of $r$ and $r'$ for models a, b, c and observables $\hmo_\mr{m}$ and $\hmo_\mr{c}$.
The figures show that the ratios do not depend on energy especially in the middle of the spectrum.
These values agree well with the predictions of random matrix theory in Tables~\ref{table1} and~\ref{table2}.

\begin{figure}
\begin{center}
\includegraphics[width=\linewidth]{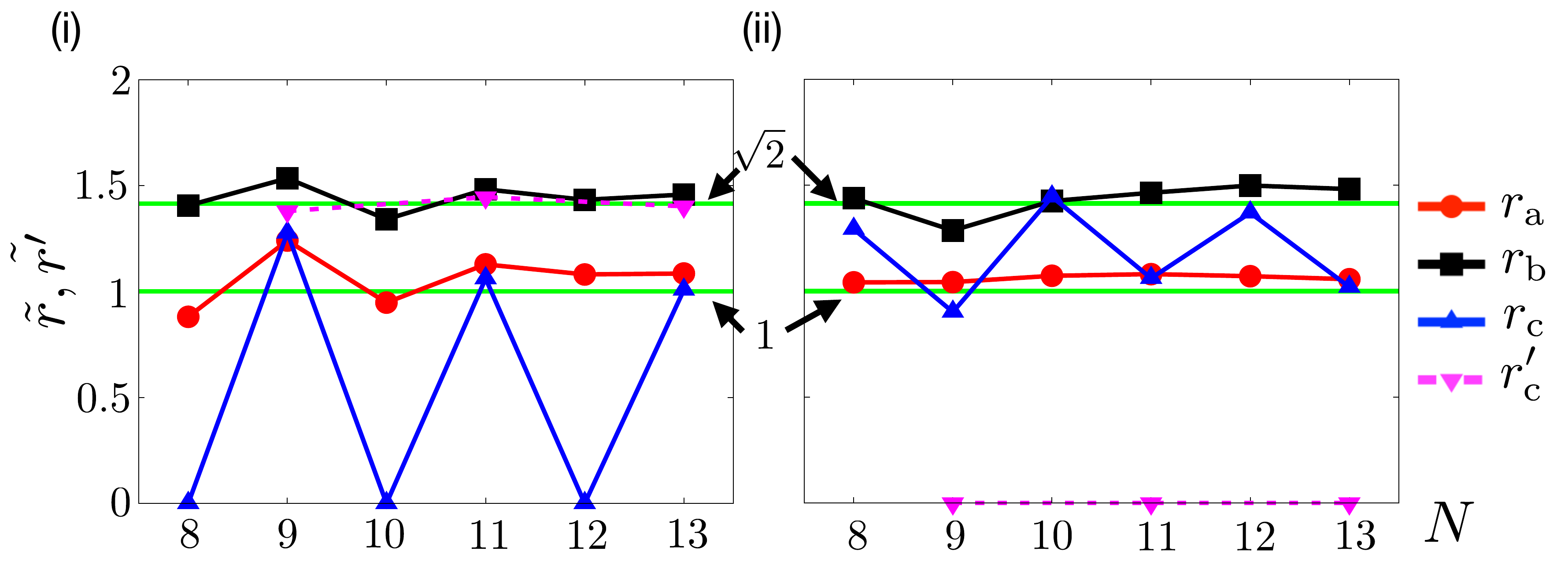}
\caption{
System-size dependences of $\tilde{r}$ and $\tilde{r}'$ for models a, b, and c with observables (i) $\hmo_\mr{m}$ and (ii) $\hmo_\mr{c}$, where the averages are obtained from the middle one-third  of the spectrum.
The results agree well with the predictions  in Tables~\ref{table1} and~\ref{table2}.
The ratio $\tilde{r}$ is $1$ and $\sqrt{2}$ for models a and b, respectively,  independent of $N$,  since model a belongs to class A and model b belongs to class AI irrespective of $N$.
On the other hand, for model c, $\tilde{r}$ and $\tilde{r'}$ depend on the parity of $N$ because it belongs to class AI for even $N$ and class AII for odd $N$.
We take  $\omega_s=(\max_\alpha E_\alpha-\min_\alpha E_\alpha)/(6N)$  for
$N\leq 10$ and $\omega_s=(\max_\alpha E_\alpha-\min_\alpha E_\alpha)/(12N)$  for $N\geq 11$.
}
\label{fig3}
\end{center}
\end{figure}

Figure~\ref{fig3} shows the average values of $r$ and $r'$ in the middle of the spectrum (denoted as $\tilde{r}$ and $\tilde{r}'$) for different system sizes.
While fluctuations such as $\Delta\mc{O}_\mr{d}$ decay exponentially with increasing the system size, the ratios $r$ and $r'$ are universal especially for $N\geq 10$.
In particular, the ratios do not depend on $N$ for models a and b, and only depend on the parity of $N$ for model c.
These universal values agree well with the predictions in Tables~\ref{table1} and~\ref{table2}.

Here, let us briefly discuss the energy scale $\Delta E_\mr{Univ}$, within which
the ratios exhibit the universal random-matrix behavior (i.e., the maximal value of $\omega_s$ with which we obtain the universal results).
Our local observables are expected to decay diffusively  with the timescale $\Delta E_\mr{Th}^{-1}\propto N^2$, where $\Delta E_\mr{Th}$ is the so-called (many-body) Thouless energy~\cite{DAlessio16}.
On the other hand, it has recently been argued~\cite{Dymarsky18} that the energy scale for the universal matrix elements obeys $\Delta E_\mr{Univ}\propto N^{-3}$ for diffusive observables.
In other words, $\Delta E_\mr{Univ}$ is smaller than the Thouless energy and the Breit-Wigner width~\cite{breit-Master} for large $N$, which are typical energy scales for the appearance of the universality for the spectral rigidity and the level number variance~\cite{Haake}.

\subsection{\label{subsec:few}Numerical results for many-body observables}
While we have verified the universal ratios for few-body operators above, we obtain similar results for many-body operators.
To demonstrate this, we introduce $l$-body spin correlations defined by
\aln{\label{oiyo}
\hmo_l:=\prod_{n=1}^l\hat{\sigma}_n^z \:\:\:(3\leq l\leq N-1).
}
We note that the case $l=N$ is omitted because it has large errors, which may be attributed to small but residual reflection symmetry of sites $i\leftrightarrow N-i$.

\begin{figure}
\begin{center}
\includegraphics[width=\linewidth]{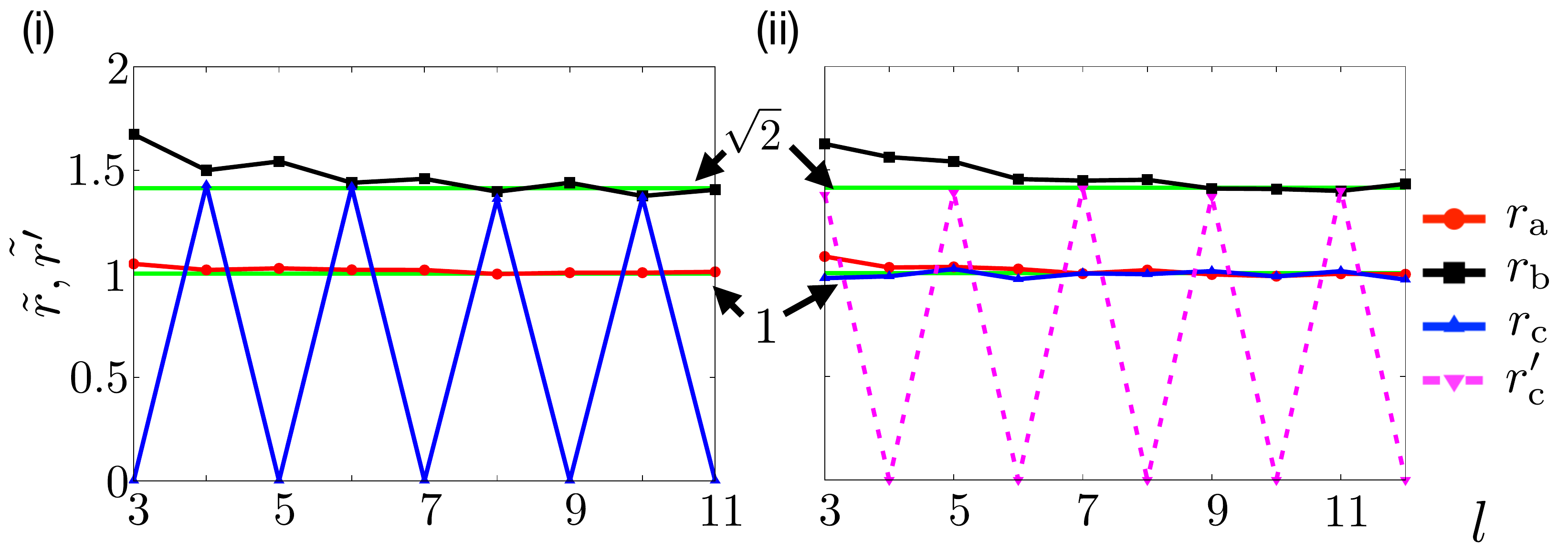}
\caption{
Dependences of $\tilde{r}$ and $\tilde{r}'$ on $l$ for models a, b, and c for (i) $N=12$ and (ii) $N=13$, where the averages are taken over the middle one-third of the spectrum.
The numerical results agree well with the predictions  in Tables~\ref{table1} and~\ref{table2}.
The ratio $\tilde{r}$ is $1$ and $\sqrt{2}$ for models a and b, respectively, independent of $N$ since model a belongs to class A and model b belongs to class AI irrespective of $N$, and $\hmo_l$ is always even.
On the other hand, for model c, $\tilde{r}$ and $\tilde{r}'$ depend on the parity of $N$ and that of $l$ because the symmetry depends on them.
We take $\omega_s=(\max_\alpha E_\alpha-\min_\alpha E_\alpha)/(12N)$.
}
\label{fig4}
\end{center}
\end{figure}

Figure~\ref{fig4} shows that $\tilde{r}$ and $\tilde{r}'$ for both even and odd $N$ are universal.
In particular, the ratios do not depend on $l$ for models a and b, and only depend on the parity of $l$ for model c.
The results agree well with the predictions  in Tables~\ref{table1} and~\ref{table2}.

It is interesting to note that the predictions in Tables~\ref{table1} and~\ref{table2} seem to agree better with the numerical data in Fig.~\ref{fig4} for larger $l$, i.e., many-body correlations.
This can be attributed to the fact that $\mc{A}(E)$ and $f(E,\omega)$ in \EQ{sred} are less dependent on $E$ and $\omega$ for many-body operators, as discussed in Ref.~\cite{Hamazaki18A}.
In fact, the approximated equality in \EQ{approximate}, which works sufficiently well for $\omega_s$ smaller than $\Delta E_\mr{Univ}$, holds better for many-body observables with larger $l$ because $\Delta E_\mr{Univ}$ becomes larger~\cite{exc-Master}.

\section{\label{sec:quench}Universality in quench dynamics in small systems}
In this section, we discuss how our universal ratio found in Sec.~\ref{sec:ratio} is evaluated from the long-time thermalization dynamics in small isolated systems after a quench.
We show that diagonal matrix elements and off-diagonal matrix elements are related to the autocorrelation function  and the temporal fluctuation of observables, respectively.
To see this, we assume that the energy fluctuation of the initial state $\hrho_0$,
\aln{
\Delta E=\sqrt{\Tr[\hrho_0\hH^2]-\Tr[\hrho_0\hH]^2},
}
is so small that the variations of $\mc{A}(E)$ and $f(E,\omega)$ in \EQ{sred} within the energy range $[E-\Delta E,E+\Delta E]$ are much smaller than the fluctuations $\Delta\mc{O}_\mr{d}$, $\Delta\mc{O}_\mr{od}$, and $\Delta\mc{O}_\mr{K}$ ($\hH$ is the Hamiltonian after the quench).
Since $\Delta E$ typically becomes larger and $\Delta\mc{O}_\mr{d}$, $\Delta\mc{O}_\mr{od}$, and $\Delta\mc{O}_\mr{K}$ become smaller with increasing the system size, our discussion is applicable to relatively small systems.
Moreover, our theory holds better for many-body correlations with larger $l$, which are in fact measurable quantities in small systems~\cite{Islam15,Kaufman16}.
This is because $\mc{A}(E)$ and $f(E,\omega)$ are expected to be less dependent on their arguments for many-body operators, as discussed in Sec.~\ref{sec:ratio}.
We also assume that the initial state is a pure state $\ket{\psi_0}$, i.e., $\hrho_0=\ket{\psi_0}\bra{\psi_0}$.

First, the standard deviation of diagonal matrix elements $\Delta\mc{O}_\mr{d}$ is related to the long-time average of the autocorrelation function~\cite{Feingold86}:
\aln{
\tilde{\mc{S}}:=\lim_{T,T'\ra\infty}\frac{1}{TT'}\int_0^T dt \int_0^{T'} dt' \braket{\psi_0|\hat{\mc{O}}(t+t')\hat{\mc{O}}(t)|\psi_0},
}
where $\hat{\mc{O}}(t):=e^{i\hat{H}t}\hat{\mc{O}}e^{-i\hat{H}t}$.
In fact, if we assume non-degeneracy (which corresponds to classes A and AI), we obtain
\aln{
\tilde{\mc{S}}= \sum_\alpha |c_\alpha|^2{\mc{O}}_{\alpha\alpha}^2,
}
where $c_\alpha:=\braket{E_\alpha|\psi_0}$.
Moreover, the long-time average of the expectation value of $\hmo$ is
\aln{
\tilde{{\mc{O}}}:=\lim_{T\ra\infty}\frac{1}{T}\int_0^T dt\braket{\psi_0|\hat{\mc{O}}(t)|\psi_0}=\sum_\alpha |c_\alpha|^2 \mc{O}_{\alpha\alpha}.
}
Thus, we obtain
\aln{\label{bunsan}
\tilde{\mc{S}}-\tilde{{\mc{O}}}^2=\sum_\alpha \lrl{|c_\alpha|^2\lrs{{\mc{O}}_{\alpha\alpha}-\sum_\beta |c_\beta|^2 {\mc{O}}_{\beta\beta}}^2}.
}
By assuming that $c_\alpha$ distributes unbiasedly over $\alpha$ (i.e., the effective dimension $\lrs{\sum_\alpha |c_\alpha|^4}^{-1}$~\cite{Mori18} is sufficiently large) with a sufficiently small width $\Delta E$, the right-hand side of \EQ{bunsan} is approximated by the variance of diagonal matrix elements,
\aln{
\tilde{\mc{S}}-\tilde{{\mc{O}}}^2\simeq \Delta \mc{O}_\mr{d}^2.
}

Next, the standard deviation of off-diagonal matrix elements is evaluated from the long-time average of temporal fluctuations at each time:
\aln{
\tilde{\mc{T}}^2:=\lim_{T\ra\infty}\frac{1}{T}\int_0^T dt[\braket{\psi_0|\hat{\mc{O}}(t)|\psi_0}-\tilde{\mc{O}}]^2.
}
By assuming non-degeneracy in gaps of eigenenergies, we obtain~\cite{Reimann08}
\aln{\label{off}
\tilde{\mc{T}}^2=\sum_{\alpha\neq\beta}|c_\alpha|^2|c_\beta|^2 |\mc{O}_{\alpha\beta}|^2.
}
Again, assuming that $c_\alpha$ distributes unbiasedly over $\alpha$ with a sufficiently small width $\Delta E$, the right-hand side of \EQ{off} is approximated by the variance of off-diagonal matrix elements,
\aln{
\tilde{\mc{T}}^2\simeq \Delta \mc{O}_\mr{od}^2.
}

From these discussions, we obtain
\aln{\label{gdayo}
g:=\sqrt{\frac{\tilde{\mc{S}}-\tilde{{\mc{O}}}^2}{\tilde{\mc{T}}^2}}\simeq r
}
for classes A and AI.
Thus, we expect $g_\mr{A}= 1, g_\mr{AI,even}= \sqrt{2}$, and $g_\mr{AI,odd}= 0$.
Moreover, as shown in Appendix~\ref{app3}, we obtain
\aln{\label{gdayo2}
g\simeq\sqrt{r^2+r'^2}
}
 for class AII, which suggests $g_\mr{AII,even}= 1$ and  $g_\mr{AII,odd}= \sqrt{3}$.
The relations in Eqs.~\eqref{gdayo} and ~\eqref{gdayo2} suggest that two different fluctuations, i.e., the correlation between two distant times and the fluctuation at each time are proportional to each other with a universal ratio.
This universality cannot be described by statistical mechanics because these  fluctuations only matter in small isolated systems.
In fact, these fluctuations vanish much more rapidly than thermal fluctuations described by statistical mechanics.
Note that this relation is different from the well-known relations about autocorrelation functions, such as the Wiener-Khintchine theorem.
In fact,  $\mc{T}^2$ is different from the squared power spectrum.

\begin{figure}
\begin{center}
\includegraphics[width=7cm]{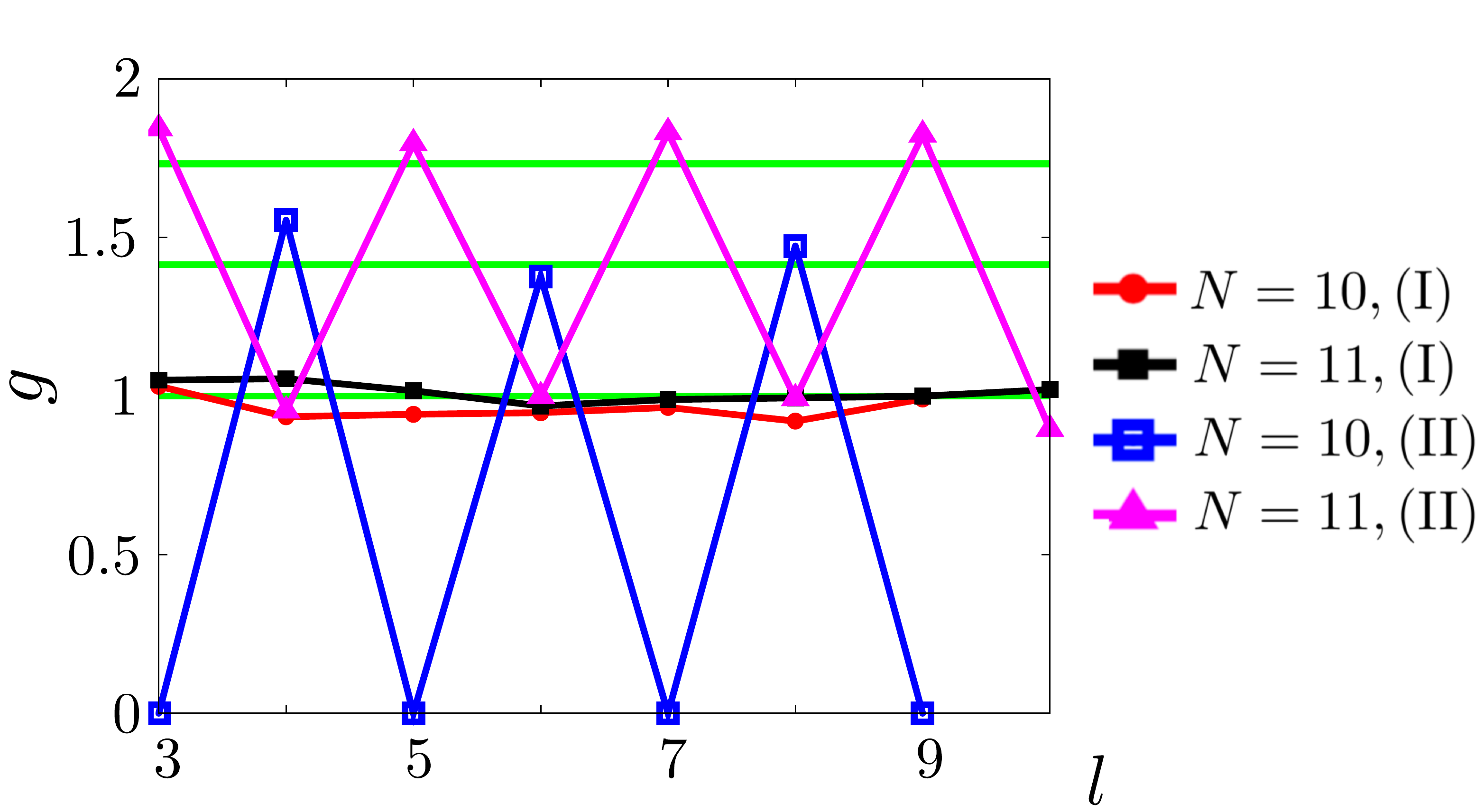}
\caption{
Dependence of $g$ on $l$ for different system sizes ($N=10,11$) and different quenches (cases (I) and (II)).
For all these cases, the results agree well with the predictions of random matrices in Table~\ref{gatai}.
The results show the averages over different initial states,  which are energy eigenstates of the initial Hamiltonian taken from the middle one-fourth of the spectrum.
}
\label{fig5}
\end{center}
\end{figure}

We demonstrate the validity of Eqs.~\eqref{gdayo} and~\eqref{gdayo2} by considering two types of the quench of our models a and c.
For the first case, which we refer to as case (I), we consider an initial Hamiltonian in \EQ{spinchain} with
$h=0.5,h'=-1.05$ and $D=1$ to model a with $h=0.5,h'=-1.05$ and $D=0.9$.
For the second case, which we refer to as case (II), we consider an initial Hamiltonian in \EQ{spinchain} with
$h=0.03,h'=-0.063$ and $D=0.9$ to model c with $h=h'=0$ and $D=0.9$.
For both cases, the initial state is chosen to be a highly excited eigenstate of the initial Hamiltonian.
Figure~\ref{fig5} shows the factor $g$ for these quenches by taking the $l$-body correlations $\hmo_l$.
For both of the quenches and all $N$ and $l$, $g$ agrees well with the predictions of random matrix theory, which  is summarized in Table~\ref{gatai}.

\begin{table}
  \begin{center}
    \caption{Values of $g$ predicted by random matrix theory. We show the results for different  quenches, different observables $\hmo_l$, and different system sizes $N$.}\label{gatai}
    \begin{tabular}{ccccc} \hline\hline
    Quench&  (I)  & (I) & (II)  & (II)\\
    Size&   $N=\mr{even}$ & $N=\mr{odd}$ & $N=\mr{even}$ & $N=\mr{odd}$\\ \hline
        Symmetry&  A  & A & AI  & AII\\ \hline

       $\hmo_l$ ($l=\mr{even}$)  &  1 & 1 & $\sqrt{2}$ & 1 \\
     $\hmo_l$ ($l=\mr{odd}$) &  1 & 1 & 0 & $\sqrt{3}$\\  \hline\hline
    \end{tabular}
  \end{center}
\end{table}

\section{\label{sec:outro}Conclusions and Outlook}
We have introduced a one-dimensional nonintegrable spin model that covers Dyson's three symmetries by varying the strengths of transverse, longitudinal fields and the DM interaction.
We have analyzed nearest-neighbor spacing distributions of the models with different symmetry and shown that they obey those of random matrices with the corresponding symmetry (GUE, GOE, and GSE).
We have also shown that the crossover transitions of the distributions occur when we slowly change the symmetry by varying parameters.
We have studied the ratios between the standard deviations of diagonal and off-diagonal matrix elements, which become universal values that depend only on symmetries of the Hamiltonian and the observable.
We have demonstrated the universality by using our nonintegrable models in addition to the predictions by random matrix theory.
Finally, we have discussed that these ratios are evaluated from long-time dynamics of small isolated quantum systems.

Our nonintegrable model provides an excellent playground to investigate spectral transitions between Dyson's three different symmetries.
As shown in Section~\ref{sec:level}, our model exhibits a crossover transition of nearest-neighbor spacing distributions by varying parameters  for a finite system size.
It is an interesting future challenge to understand how this transition becomes sharper with increasing the system size towards the thermodynamic limit.
While such a question was discussed in random matrices~\cite{Guhr98}, it is still  nontrivial for nonintegrable many-body systems with local interactions.
%It is also interesting to investigate how the change of symmetry leads to different universal consequences in our nonintegrable many-body models other than the factor $g$ discussed in our work.
%We leave these questions for future investigation.

\begin{acknowledgements}
\emph{Acknowledgments.---}
We are grateful to Kazuya Fujimoto, Takashi Mori, Takahiro Sagawa, Akira Shimizu, and Kazue Kudo for valuable comments.
This work was supported by
KAKENHI Grant No. JP18H01145 and
a Grant-in-Aid for Scientific Research on Innovative Areas ``Topological Materials Science" (KAKENHI Grant No. JP15H05855)
from the Japan Society for the Promotion of Science.
R. H. was supported by the Japan Society for the Promotion of Science through Program for Leading Graduate Schools (ALPS) and JSPS fellowship (JSPS KAKENHI Grant No. JP17J03189).

\end{acknowledgements}

%\begin{widetext}

\appendix

\section{\label{app1}Another indicator of level-spacing distributions}
References~\cite{Atas13,Atas13J} investigated the ratio of consecutive level spacings of random matrices defined by
\aln{
R=\frac{\min(E_{\alpha+1}-E_\alpha,E_{\alpha}-E_{\alpha-1})}{\max(E_{\alpha+1}-E_\alpha,E_{\alpha}-E_{\alpha-1})}.
}
This quantity does not require the unfolding procedure in contrast with $P(s)$ discussed in the main text.

\begin{figure*}
\begin{center}
\includegraphics[width=14cm]{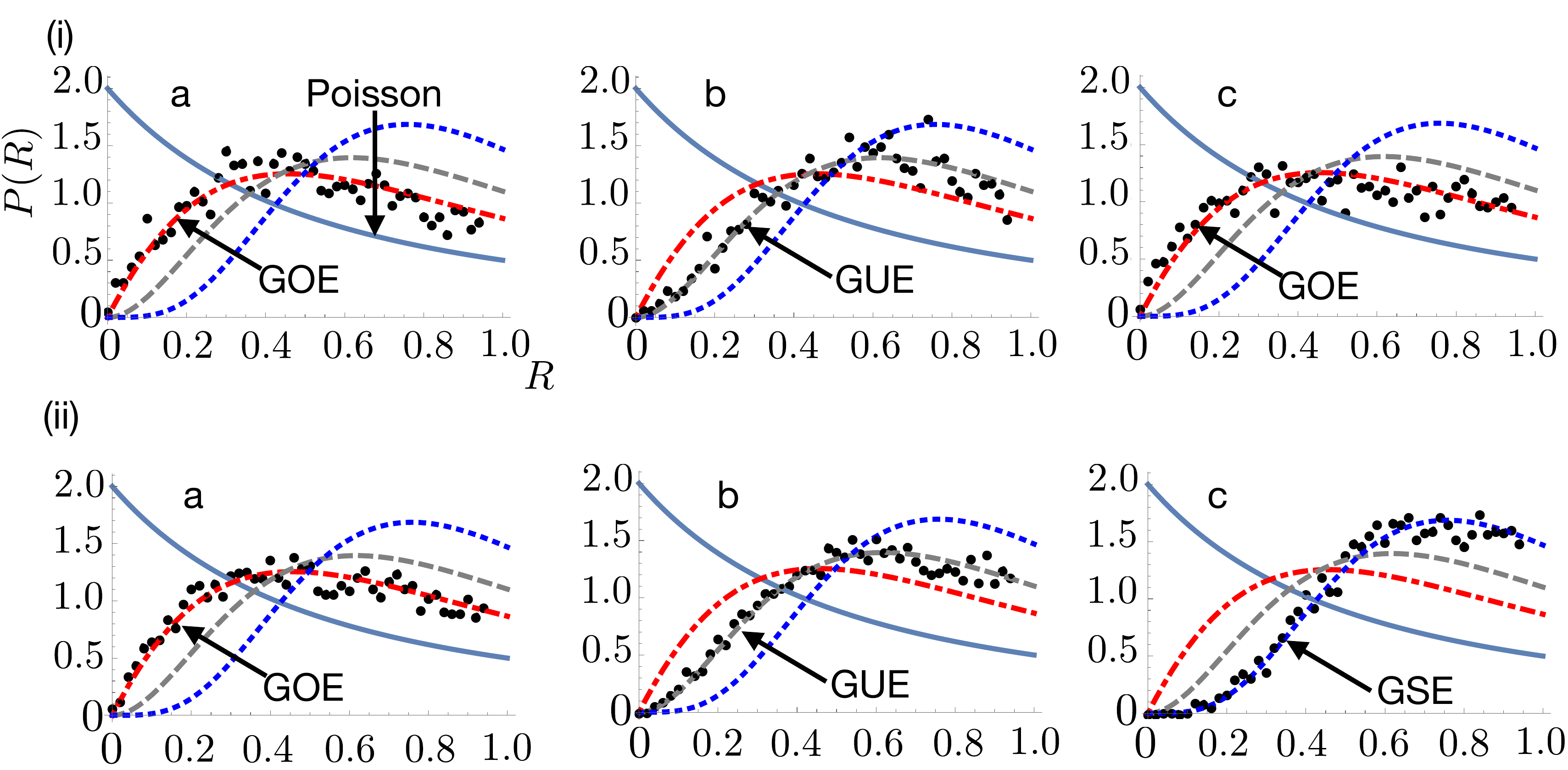}
\caption{
Probability distribution of $R$ for different models (left) a, (middle) b, and (right) c with different system sizes (i) $N=12$ and (ii) $N=13$.
Each result shown by the large black dots is well described by the prediction of random-matrix ensemble with the corresponding symmetry, i.e., GUE (dashed), GOE (dot-dashed), or GSE (dotted) in Eq.~\eqref{eq:rlevel}, rather than the case of the Poisson distribution (solid) in Eq.~\eqref{eq:rlevel2}.
}
\label{fig8}
\end{center}
\end{figure*}

Figure~\ref{fig8} shows the probability distribution of $R$ for different models a, b, and c with different system sizes.
Each result is well described by the prediction of random-matrix ensemble with the corresponding symmetry, i.e., GUE, GOE, or GSE.
Note that the random-matrix predictions discussed in Ref.~\cite{Atas13} are
\aln{\label{eq:rlevel}
P(R)&=\frac{2}{Z_\beta}\frac{(R+R^2)^\beta}{(1+R+R^2)^{1+3\beta/2}}\Theta(1-R),
}
where $\Theta(x)$ is the Heaviside unit-step function and $\beta=1,2,$ and 4 for GOE, GUE, and GSE, respectively.
The constants $Z_\beta$ are $Z_1=8/27$, $Z_2=4\pi/(81\sqrt{3})$, and $Z_4=4\pi/(729\sqrt{3})$.
Moreover, we have
\aln{\label{eq:rlevel2}
P(R)=\frac{2}{(1+R)^2}\Theta(1-R)
}
for the Poisson distribution.

\section{\label{app2}Derivations of the universal ratios $r$ and $r'$ by random matrix theory}
Here we derive the universal ratios from random matrix theory.
While the ratios for classes A (GUE) and AI (GOE) were discussed in Refs.~\cite{DAlessio16,Mondaini17}, their calculations are not accurate because they ignore non-negligible correlations for random vectors.

\subsection{Class A: calculation by GUE}
Let us examine class A.
We calculate the ensemble average of diagonal and off-diagonal matrix elements of an observable $\hmo$, which can be diagonalized as
\aln{
\hmo=\sum_{i=1}^d\ket{a_i}\bra{a_i},
}
where $a_i$ and $\ket{a_i}$ are eigenvalues and eigenstates for the observable, and $d$ is the dimension of the matrix.
The matrix elements can be written as
\aln{
{\mc{O}}_{\alpha\beta}=\sum_ia_iU_{\alpha i}U_{i\beta},
}
where $U_{\alpha i}:=\braket{E_\alpha|a_i}$ denotes a transformation function between bases $\ket{E_\alpha}$ and $\ket{a_i}$.
Let us assume that the Hamiltonian possesses the symmetry of GUE and that an  observable is fixed.
Then, it is known that $U$ is distributed uniformly with respect to the unitary Haar measure~\cite{Haake}.
In this case, we have the following moments of $U$~\cite{Brody81}:
\aln{
\av{U_{\alpha i}U_{i\beta}}&=\frac{1}{d}\delta_{\alpha\beta},\\
\av{|U_{\alpha i}|^2|U_{i\beta}|^2}&=\frac{1+\delta_{\alpha\beta}}{d(d+1)},\\
\av{|U_{\alpha i}|^2|U_{j\alpha}|^2}&=\frac{1+\delta_{ij}}{d(d+1)},\\
\av{U_{\alpha i}U_{i\beta}U_{\beta j}U_{j\alpha}}&=-\frac{1}{d(d-1)(d+1)}\:\:\:(\alpha\neq\beta,i\neq j)\label{highyo},
}
where $\av{\cdots}$ denotes the ensemble average (the average with respect to the unitary Haar measure).
These lead to the following averages and the variances of the matrix elements:
\aln{\label{sim}
\av{\mc{O}_{\alpha\beta}}&=\frac{\delta_{\alpha\beta}}{d}\sum_ia_i,\\
\av{\mc{O}_{\alpha\alpha}^2}-\av{\mc{O}_{\alpha\alpha}}^2&=\sum_{ij}a_ia_j\av{|U_{\alpha i}|^2|U_{j\alpha}|^2}-\lrs{\frac{1}{d}\sum_ia_i}^2\NON
&=\frac{1}{d(d+1)}\sum_ia_i^2+\frac{1}{d(d+1)}\lrs{\sum_{i}a_i}^2\NON
&\:\:\:-\lrs{\frac{1}{d}\sum_ia_i}^2\NON
&=\frac{1}{d+1}\lrl{\frac{1}{d}\sum_ia_i^2-\lrs{\frac{1}{d}\sum_ia_i}^2},\label{sim2}
}
and
\aln{\label{sim3}
\av{|\mc{O}_{\alpha\beta}|^2}&=\sum_{ij}a_ia_j\av{U_{\alpha i}U_{i\beta}U_{\beta j}U_{j\alpha}}\NON
&=\frac{1}{d(d+1)}\sum_ia_i^2-\frac{1}{d(d-1)(d+1)}\sum_{i\neq j}a_ia_j\NON
&=\frac{d}{(d+1)(d-1)}\lrl{\frac{1}{d}\sum_ia_i^2-\lrs{\frac{1}{d}\sum_ia_i}^2}
}
for $\alpha\neq\beta$.
We note that in Refs.~\cite{DAlessio16,Mondaini17} the term $-\lrs{\frac{1}{d}\sum_ia_i}^2$ is missing in the variances~\eqref{sim2}  and~\eqref{sim3} because the authors of Refs.~\cite{DAlessio16,Mondaini17} ignore some correlations such as \EQ{highyo}, which contribute to the lowest-order term after the summation over $i$ and $j$.

From Eqs.~\eqref{sim2}  and~\eqref{sim3}, we obtain the  ratio
\aln{
r_\mr{A}=\sqrt{\frac{d-1}{d}},
}
which becomes 1 for large $d$.

\subsection{Class AI: calculation by GOE}
Next, we consider the Hamiltonian that belongs to GOE and the observables that are even under $\hat{T}$.
We assume that neither the Hamiltonian nor the observable has a degeneracy.
For
$\hmo=\sum_{i=1}^{d}a_i\ket{a_i}\bra{a_i}$,
 we can assume that $\hat{T}\ket{a_i}=\ket{a_i}$ and $\hat{T}\ket{E_\alpha}=\ket{E_\alpha}$ without loss of generality.
To see this, let us consider $\ket{E_\alpha}$ as an example. Since $\hH\hat{T}\ket{E_\alpha}=\hat{T}\hH\ket{E_\alpha}=E_\alpha\hat{T}\ket{E_\alpha}$ and no degeneracy exists, we obtain $\hat{T}\ket{E_\alpha}=e^{i\theta}\ket{E_\alpha}$ for some $\theta\in [0,2\pi)$. If we redefine the eigenstate as $\ket{E_\alpha'}:=e^{i\theta/2}\ket{E_\alpha}$,  we obtain $\hat{T}\ket{E_\alpha'}=\ket{E_\alpha'}$ as desired.
Now, the matrix elements can be taken to be real because
\aln{
\mc{O}_{\alpha\beta}&=(\ket{E_\alpha},\hat{\mc{O}}\ket{E_\beta})\NON
&=(\hat{T}\ket{E_\alpha},\hat{T}\hat{\mc{O}}\hat{T}^{-1}\hat{T}\ket{E_\beta})^*\NON
&=(\ket{E_\alpha},(+\hat{\mc{O}})\ket{E_\beta})^*=\mc{O}_{\alpha\beta}^*,
}
where $(\vec{a},\vec{b})$ denotes an inner product of $\vec{a}$ and $\vec{b}$ and we have used $(\hat{T}\vec{a},\hat{T}\vec{b})^*=(\vec{a},\vec{b})$.

In this case,  the basis transformation $U_{\alpha i}:=\braket{E_\alpha|a_i}$ is distributed uniformly with respect to the \textit{orthogonal} Haar measure, which reflects the fact that $U_{\alpha i}$ can be taken as being real: $\braket{E_\alpha|a_i}=(\hat{T}\ket{E_\alpha},\hat{T}\ket{a_i})^*=\braket{E_\alpha|a_i}^*$.
Then, the moments of $U$ can be written as
\aln{
\av{U_{\alpha i}U_{i\beta}}&=\frac{1}{d}\delta_{\alpha\beta},\label{highO0}\\
\av{U_{\alpha i}^2U_{i\beta}^2}&=\frac{1+2\delta_{\alpha\beta}}{d(d+2)},\\
\av{U_{\alpha i}^2U_{j\alpha}^2}&=\frac{1+2\delta_{ij}}{d(d+2)},\\
\av{U_{\alpha i}U_{i\beta}U_{\beta j}U_{j\alpha}}&=-\frac{1}{d(d-1)(d+2)}\:\:\:(\alpha\neq\beta,i\neq j)\label{highO}.
}
Using Eqs. (\ref{highO0})-(\ref{highO}), the averages and the variances of diagonal and off-diagonal matrix elements can be calculated as
\aln{
\av{\mc{O}_{\alpha\beta}}&=\frac{\delta_{\alpha\beta}}{d}\sum_ia_i,\\
\av{(\mc{O}_{\alpha\alpha}-\av{\mc{O}_{\alpha\alpha}})^2}&=\frac{2}{d+2}\lrl{\frac{1}{d}\sum_ia_i^2-\lrs{\frac{1}{d}\sum_ia_i}^2},
}
and
\aln{
\av{\mc{O}_{\alpha\beta}^2}=\frac{d}{(d+2)(d-1)}\lrl{\frac{1}{d}\sum_ia_i^2-\lrs{\frac{1}{d}\sum_ia_i}^2}
}
for $\alpha\neq\beta$,
which can be obtained with calculations similar to GUE.
Therefore, we obtain
\aln{
r_\mr{GOE, even}=\sqrt{\frac{2(d-1)}{d}},
}
which becomes $\sqrt{2}$ for large $d$.

Next, if the Hamiltonian belongs to GOE and the observable is odd under $\hat{T}$, $r_\mr{GOE,odd}=0$ is obtained.
This result arises from the vanishing diagonal matrix elements:
\aln{\mc{O}_{\alpha\alpha}&=(\hat{T}\ket{E_\alpha},\hat{T}\hmo\hat{T}^{-1}\hat{T}\ket{E_\alpha})^*\NON
&=(\ket{E_\alpha},(-\hmo)\ket{E_\alpha})^*\NON
&=-\mc{O}_{\alpha\alpha}\NON
&= 0.
}

\subsection{Class AII: calculation by GSE}
If the Hamiltonian belongs to GSE, it can be written as $\hH=\sum_{\alpha=1}^{d/2}E_\alpha(\ket{E_\alpha}\bra{E_\alpha}+\ket{\tilde{E_\alpha}}\bra{\tilde{E_\alpha}})$.
Here, we note that $d$ is always even in GSE.
Let us first consider that the observable is even under $\hat{T}$.
In this case, the observable can be written as $\hmo=\sum_{i=1}^da_i\ket{a_i}\bra{a_i}=\sum_{i'=1}^{d/2}a_{i'}(\ket{a_{i'}}\bra{a_{i'}}+\ket{\tilde{a_{i'}}}\bra{\tilde{a_{i'}}})$, where $\ket{\tilde{a_{i'}}}:=\hat{T}\ket{a_{i'}}$ and $\hmo\ket{\tilde{a_{i'}}}=a_{i'}\ket{\tilde{a_{i'}}}$.
We assume that the Hamiltonian and the observable have no degeneracy except for Kramers degeneracies.
By calculating (higher) moments of $\braket{E_\alpha|a_{i'}}$ and $\braket{E_\alpha|\tilde{a_{i'}}}$ using random matrix theory, we obtain the averages and the variances of diagonal and off-diagonal matrix elements as follows~\cite{HamazakiM}:
\aln{\label{gseE}
\av{\mc{O}_{\alpha\beta}}&=\frac{\delta_{\alpha\beta}}{d}\sum_ia_i,\\
\av{(\mc{O}_{\alpha\alpha}-\av{\mc{O}_{\alpha\alpha}})^2}&=\frac{1}{d+1}\lrl{\frac{1}{d}\sum_ia_i^2-\lrs{\frac{1}{d}\sum_ia_i}^2},\\
\av{|\mc{O}_{\alpha\beta}|^2}
&=\frac{d}{(d-2)(d+1)}\lrl{\frac{1}{d}\sum_ia_i^2-\lrs{\frac{1}{d}\sum_ia_i}^2}
}
for $\alpha\neq\beta$.
 Thus we obtain
\aln{
r_\mr{GSE,even}= \sqrt{\frac{d-2}{d}},
}
which becomes 1 for large $d$.
For the ratio concerning the Kramers pair, we obtain $r'_\mr{GSE,even}=0$, which results from the vanishing $\mc{O}_{\alpha\tilde{\alpha}}$:
\aln{\mc{O}_{\alpha\tilde{\alpha}}&=(\hat{T}\ket{E_\alpha},\hat{T}\hmo\hat{T}^{-1}\hat{T}\hat{T}\ket{E_\alpha})^*\NON
&=(\ket{\tilde{E_\alpha}},\hmo(-1)\ket{E_\alpha})^*\NON
&=-\mc{O}_{\alpha\tilde{\alpha}}\NON
&= 0.
}
Here we have used $\hat{T}^2=-1$.

Finally, we consider the case in which the Hamiltonian belongs to GSE and the observable is odd under $\hat{T}$.
In this case, the observable can be written as $\hmo=\sum_{i=1}^da_i\ket{a_i}\bra{a_i}=\sum_{a_{i'}>0}a_{i'}(\ket{a_{i'}}\bra{a_{i'}}-\ket{\tilde{a_{i'}}}\bra{\tilde{a_{i'}}})$, where $\ket{\tilde{a_{i'}}}:=\hat{T}\ket{a_{i'}}$ and $\hmo\ket{\tilde{a_{i'}}}=-a_{i'}\ket{\tilde{a_{i'}}}$.
By calculating the (higher) moments of inner products such as $\braket{E_\alpha|a_{i'}}$ and $\braket{E_\alpha|\tilde{a_{i'}}}$ using random matrix theory, we obtain the averages and the variances of matrix elements~\cite{HamazakiM}.
First, the averages of the matrix elements are all zero since $\sum_ia_i=0$:
\aln{\label{gseE2}
\av{\mc{O}_{\alpha\beta}}=0.
}
For the variances, we obtain
\aln{
\av{(\mc{O}_{\alpha\alpha})^2}&=\frac{1}{d+1}\lrl{\frac{1}{d}\sum_ia_i^2},\\
\av{|\mc{O}_{\alpha\beta}|^2}
&=\frac{1}{d+1}\lrl{\frac{1}{d}\sum_ia_i^2}\:\:\:(E_\alpha\neq E_\beta),\\
\av{|\mc{O}_{\alpha\tilde{\alpha}}|^2}&=\frac{2}{d+1}\lrl{\frac{1}{d}\sum_ia_i^2}.
}
We then obtain
\aln{
r_\mr{GSE,odd}&= 1,\\
r'_\mr{GSE,odd}&=\sqrt{2}.\label{gseE3}
}

\section{\label{app3}Factor $g$ in the presence of the Kramers degeneracy}
In this appendix, we derive \EQ{gdayo2}.
We first expand the initial state as
\aln{
\ket{\psi_0}=\sum_\alpha (c_\alpha\ket{E_\alpha}+c_{\tilde{\alpha}}\ket{\tilde{E_\alpha}})=\sum_\alpha \sqrt{p_\alpha}\ket{\phi_\alpha},
}
where
\aln{
p_\alpha=|c_\alpha|^2+|c_{\tilde{\alpha}}|^2
}
and
\aln{
 \ket{\phi_\alpha}=\frac{1}{\sqrt{p_\alpha}} (c_\alpha\ket{E_\alpha}+c_{\tilde{\alpha}}\ket{\tilde{E_\alpha}}).
}

The autocorrelation function can be written as
\aln{
&\braket{\psi_0|\hat{\mc{O}}(t+t')\hat{\mc{O}}(t)|\psi_0}\NON
=& \sum_{\alpha\beta}\sqrt{p_\alpha p_\beta}
e^{iE_\alpha (t+t')}e^{-iE_\beta t}\braket{\phi_\alpha |\mc{\hat{O}}e^{-iHt'}\mc{\hat{O}}|\phi_\beta}\nonumber\\
=&
\sum_{\alpha\beta\gamma}\sqrt{p_\alpha p_\beta}
e^{i(E_\alpha-E_\beta) t-i(E_\gamma -E_\alpha)t'}\nonumber\\
&\times\lrs{\braket{\phi_\alpha |\mc{\hat{O}}|\phi_\gamma}\braket{\phi_\gamma |\mc{\hat{O}}|\phi_\beta}+\braket{\phi_\alpha |\mc{\hat{O}}|\tilde{\phi_\gamma}}\braket{\tilde{\phi_\gamma} |\mc{\hat{O}}|\phi_\beta}},
}
which leads to
\aln{
\tilde{\mc{S}}=\sum_\alpha p_\alpha\lrs{|\braket{\phi_\alpha |\mc{\hat{O}}|\phi_\alpha}|^2+|\braket{\phi_\alpha |\mc{\hat{O}}|\tilde{\phi_\alpha}}|^2}
}
by assuming non-degeneracy except for Kramers degeneracy.
Here $\ket{\tilde{\phi_\alpha}}=\hat{T}\ket{\phi_\alpha}$.
Similarly,
\aln{
\tilde{\mc{O}}=\sum_\alpha p_\alpha \braket{\phi_\alpha|\hat{\mc{O}}|\phi_\alpha}
}
and then
\aln{\label{ggse}
\tilde{\mc{S}}-\tilde{\mc{O}}^2=&\sum_\alpha \lrl{p_\alpha \lrs{\braket{\phi_\alpha|\hat{\mc{O}}|\phi_\alpha}-\sum_\alpha p_\beta \braket{\phi_\beta|\hat{\mc{O}}|\phi_\beta}}^2}\NON
&+\sum_\alpha p_\alpha|\braket{\phi_\alpha |\mc{\hat{O}}|\tilde{\phi}_\alpha}|^2.
}

Since the random-matrix-theory treatment of GSE discussed in Appendix~\ref{app2} is applicable to a general rotation in the two-dimensional Kramers-pair space, we can replace the role of $\ket{E_\alpha}$ and $\ket{\tilde{E_\alpha}}$ with $\ket{{\phi_\alpha}}$ and $\ket{\tilde{\phi_\alpha}}$ in considering the statistics of matrix elements.
Thus, the first and second terms on the right-hand side of \EQ{ggse} are approximated by $\Delta\mc{O}_\mr{d}^2$ and $\Delta\mc{O}_\mr{K}^2$, respectively, in analogy with the discussion in the main text.
We then have
\aln{
\tilde{\mc{S}}-\tilde{\mc{O}}^2\simeq \Delta\mc{O}_\mr{d}^2+\Delta\mc{O}_\mr{K}^2.
}

Next, for off-diagonal matrix elements we obtain
\aln{
\tilde{\mc{T}}^2=\sum_{\alpha\neq\beta}p_\alpha p_\beta |\braket{\phi_\alpha|\hat{\mc{O}}|\phi_\beta}|^2.
}
Again, we can safely apply the discussion in Appendix~\ref{app2} on $\ket{E_\alpha}$ and $\ket{\tilde{E_\alpha}}$ to $\ket{{\phi_\alpha}}$ and $\ket{\tilde{\phi_\alpha}}$ in considering the statistics of matrix elements.
Then, in analogy with the main text, we obtain $\tilde{\mc{T}}^2\simeq \Delta\mc{O}_\mr{od}^2$.
We thus obtain
\aln{
g\simeq \sqrt{\frac{\Delta \mc{O}_\mr{d}^2+\Delta \mc{O}_\mr{k}^2}{ \Delta \mc{O}_\mr{od}^2}}=\sqrt{r^2+r'^2}
}
for class AII.

\bibliographystyle{apsrev4-1}
\bibliography{../../../../refer_them}

\begin{thebibliography}{82}%
\makeatletter
\providecommand \@ifxundefined [1]{%
 \@ifx{#1\undefined}
}%
\providecommand \@ifnum [1]{%
 \ifnum #1\expandafter \@firstoftwo
 \else \expandafter \@secondoftwo
 \fi
}%
\providecommand \@ifx [1]{%
 \ifx #1\expandafter \@firstoftwo
 \else \expandafter \@secondoftwo
 \fi
}%
\providecommand \natexlab [1]{#1}%
\providecommand \enquote  [1]{``#1''}%
\providecommand \bibnamefont  [1]{#1}%
\providecommand \bibfnamefont [1]{#1}%
\providecommand \citenamefont [1]{#1}%
\providecommand \href@noop [0]{\@secondoftwo}%
\providecommand \href [0]{\begingroup \@sanitize@url \@href}%
\providecommand \@href[1]{\@@startlink{#1}\@@href}%
\providecommand \@@href[1]{\endgroup#1\@@endlink}%
\providecommand \@sanitize@url [0]{\catcode `\\12\catcode `\$12\catcode
  `\&12\catcode `\#12\catcode `\^12\catcode `\_12\catcode `\%12\relax}%
\providecommand \@@startlink[1]{}%
\providecommand \@@endlink[0]{}%
\providecommand \url  [0]{\begingroup\@sanitize@url \@url }%
\providecommand \@url [1]{\endgroup\@href {#1}{\urlprefix }}%
\providecommand \urlprefix  [0]{URL }%
\providecommand \Eprint [0]{\href }%
\providecommand \doibase [0]{http://dx.doi.org/}%
\providecommand \selectlanguage [0]{\@gobble}%
\providecommand \bibinfo  [0]{\@secondoftwo}%
\providecommand \bibfield  [0]{\@secondoftwo}%
\providecommand \translation [1]{[#1]}%
\providecommand \BibitemOpen [0]{}%
\providecommand \bibitemStop [0]{}%
\providecommand \bibitemNoStop [0]{.\EOS\space}%
\providecommand \EOS [0]{\spacefactor3000\relax}%
\providecommand \BibitemShut  [1]{\csname bibitem#1\endcsname}%
\let\auto@bib@innerbib\@empty
%</preamble>
\bibitem [{\citenamefont {Wigner}(1951)}]{Wigner51}%
  \BibitemOpen
  \bibfield  {author} {\bibinfo {author} {\bibfnamefont {E.~P.}\ \bibnamefont
  {Wigner}},\ }in\ \href
  {https://www.cambridge.org/core/journals/mathematical-proceedings-of-the-cambridge-philosophical-society/article/on-the-statistical-distribution-of-the-widths-and-spacings-of-nuclear-resonance-levels/97EAA86F8F11C09D67D47CD700107D34}
  {\emph {\bibinfo {booktitle} {Proc. Cambridge Philos. Soc.}}},\ Vol.~\bibinfo
  {volume} {47}\ (\bibinfo {organization} {Cambridge Univ Press},\ \bibinfo
  {year} {1951})\ pp.\ \bibinfo {pages} {790--798}\BibitemShut {NoStop}%
\bibitem [{\citenamefont {Brody}\ \emph {et~al.}(1981)\citenamefont {Brody},
  \citenamefont {Flores}, \citenamefont {French}, \citenamefont {Mello},
  \citenamefont {Pandey},\ and\ \citenamefont {Wong}}]{Brody81}%
  \BibitemOpen
  \bibfield  {author} {\bibinfo {author} {\bibfnamefont {T.~A.}\ \bibnamefont
  {Brody}}, \bibinfo {author} {\bibfnamefont {J.}~\bibnamefont {Flores}},
  \bibinfo {author} {\bibfnamefont {J.~B.}\ \bibnamefont {French}}, \bibinfo
  {author} {\bibfnamefont {P.~A.}\ \bibnamefont {Mello}}, \bibinfo {author}
  {\bibfnamefont {A.}~\bibnamefont {Pandey}}, \ and\ \bibinfo {author}
  {\bibfnamefont {S.~S.~M.}\ \bibnamefont {Wong}},\ }\href {\doibase
  10.1103/RevModPhys.53.385} {\bibfield  {journal} {\bibinfo  {journal} {Rev.
  Mod. Phys.}\ }\textbf {\bibinfo {volume} {53}},\ \bibinfo {pages} {385}
  (\bibinfo {year} {1981})}\BibitemShut {NoStop}%
\bibitem [{\citenamefont {Haq}\ \emph {et~al.}(1982)\citenamefont {Haq},
  \citenamefont {Pandey},\ and\ \citenamefont {Bohigas}}]{Haq82}%
  \BibitemOpen
  \bibfield  {author} {\bibinfo {author} {\bibfnamefont {R.~U.}\ \bibnamefont
  {Haq}}, \bibinfo {author} {\bibfnamefont {A.}~\bibnamefont {Pandey}}, \ and\
  \bibinfo {author} {\bibfnamefont {O.}~\bibnamefont {Bohigas}},\ }\href
  {\doibase 10.1103/PhysRevLett.48.1086} {\bibfield  {journal} {\bibinfo
  {journal} {Phys. Rev. Lett.}\ }\textbf {\bibinfo {volume} {48}},\ \bibinfo
  {pages} {1086} (\bibinfo {year} {1982})}\BibitemShut {NoStop}%
\bibitem [{\citenamefont {Zelevinsky}\ \emph {et~al.}(1996)\citenamefont
  {Zelevinsky}, \citenamefont {Brown}, \citenamefont {Frazier},\ and\
  \citenamefont {Horoi}}]{Zelevinsky96}%
  \BibitemOpen
  \bibfield  {author} {\bibinfo {author} {\bibfnamefont {V.}~\bibnamefont
  {Zelevinsky}}, \bibinfo {author} {\bibfnamefont {B.~A.}\ \bibnamefont
  {Brown}}, \bibinfo {author} {\bibfnamefont {N.}~\bibnamefont {Frazier}}, \
  and\ \bibinfo {author} {\bibfnamefont {M.}~\bibnamefont {Horoi}},\
  }\href@noop {} {\bibfield  {journal} {\bibinfo  {journal} {Physics reports}\
  }\textbf {\bibinfo {volume} {276}},\ \bibinfo {pages} {85} (\bibinfo {year}
  {1996})}\BibitemShut {NoStop}%
\bibitem [{\citenamefont {Weidenm\"uller}\ and\ \citenamefont
  {Mitchell}(2009)}]{Weidenmuller09}%
  \BibitemOpen
  \bibfield  {author} {\bibinfo {author} {\bibfnamefont {H.~A.}\ \bibnamefont
  {Weidenm\"uller}}\ and\ \bibinfo {author} {\bibfnamefont {G.~E.}\
  \bibnamefont {Mitchell}},\ }\href {\doibase 10.1103/RevModPhys.81.539}
  {\bibfield  {journal} {\bibinfo  {journal} {Rev. Mod. Phys.}\ }\textbf
  {\bibinfo {volume} {81}},\ \bibinfo {pages} {539} (\bibinfo {year}
  {2009})}\BibitemShut {NoStop}%
\bibitem [{\citenamefont {Mitchell}\ \emph {et~al.}(2010)\citenamefont
  {Mitchell}, \citenamefont {Richter},\ and\ \citenamefont
  {Weidenm\"uller}}]{Mitchell10}%
  \BibitemOpen
  \bibfield  {author} {\bibinfo {author} {\bibfnamefont {G.~E.}\ \bibnamefont
  {Mitchell}}, \bibinfo {author} {\bibfnamefont {A.}~\bibnamefont {Richter}}, \
  and\ \bibinfo {author} {\bibfnamefont {H.~A.}\ \bibnamefont
  {Weidenm\"uller}},\ }\href {\doibase 10.1103/RevModPhys.82.2845} {\bibfield
  {journal} {\bibinfo  {journal} {Rev. Mod. Phys.}\ }\textbf {\bibinfo {volume}
  {82}},\ \bibinfo {pages} {2845} (\bibinfo {year} {2010})}\BibitemShut
  {NoStop}%
\bibitem [{\citenamefont {Rosenzweig}\ and\ \citenamefont
  {Porter}(1960)}]{Rosenzweig60}%
  \BibitemOpen
  \bibfield  {author} {\bibinfo {author} {\bibfnamefont {N.}~\bibnamefont
  {Rosenzweig}}\ and\ \bibinfo {author} {\bibfnamefont {C.~E.}\ \bibnamefont
  {Porter}},\ }\href {\doibase 10.1103/PhysRev.120.1698} {\bibfield  {journal}
  {\bibinfo  {journal} {Phys. Rev.}\ }\textbf {\bibinfo {volume} {120}},\
  \bibinfo {pages} {1698} (\bibinfo {year} {1960})}\BibitemShut {NoStop}%
\bibitem [{\citenamefont {Camarda}\ and\ \citenamefont
  {Georgopulos}(1983)}]{Camarda83}%
  \BibitemOpen
  \bibfield  {author} {\bibinfo {author} {\bibfnamefont {H.~S.}\ \bibnamefont
  {Camarda}}\ and\ \bibinfo {author} {\bibfnamefont {P.~D.}\ \bibnamefont
  {Georgopulos}},\ }\href {\doibase 10.1103/PhysRevLett.50.492} {\bibfield
  {journal} {\bibinfo  {journal} {Phys. Rev. Lett.}\ }\textbf {\bibinfo
  {volume} {50}},\ \bibinfo {pages} {492} (\bibinfo {year} {1983})}\BibitemShut
  {NoStop}%
\bibitem [{\citenamefont {Haller}\ \emph {et~al.}(1983)\citenamefont {Haller},
  \citenamefont {K{\"o}ppel},\ and\ \citenamefont {Cederbaum}}]{Haller83}%
  \BibitemOpen
  \bibfield  {author} {\bibinfo {author} {\bibfnamefont {E.}~\bibnamefont
  {Haller}}, \bibinfo {author} {\bibfnamefont {H.}~\bibnamefont {K{\"o}ppel}},
  \ and\ \bibinfo {author} {\bibfnamefont {L.}~\bibnamefont {Cederbaum}},\
  }\href {https://www.sciencedirect.com/science/article/pii/0009261483870018}
  {\bibfield  {journal} {\bibinfo  {journal} {Chemical physics letters}\
  }\textbf {\bibinfo {volume} {101}},\ \bibinfo {pages} {215} (\bibinfo {year}
  {1983})}\BibitemShut {NoStop}%
\bibitem [{\citenamefont {Abramson}\ \emph {et~al.}(1984)\citenamefont
  {Abramson}, \citenamefont {Field}, \citenamefont {Imre}, \citenamefont
  {Innes},\ and\ \citenamefont {Kinsey}}]{Abramson84}%
  \BibitemOpen
  \bibfield  {author} {\bibinfo {author} {\bibfnamefont {E.}~\bibnamefont
  {Abramson}}, \bibinfo {author} {\bibfnamefont {R.~W.}\ \bibnamefont {Field}},
  \bibinfo {author} {\bibfnamefont {D.}~\bibnamefont {Imre}}, \bibinfo {author}
  {\bibfnamefont {K.}~\bibnamefont {Innes}}, \ and\ \bibinfo {author}
  {\bibfnamefont {J.~L.}\ \bibnamefont {Kinsey}},\ }\href
  {https://aip.scitation.org/doi/abs/10.1063/1.447006} {\bibfield  {journal}
  {\bibinfo  {journal} {The Journal of chemical physics}\ }\textbf {\bibinfo
  {volume} {80}},\ \bibinfo {pages} {2298} (\bibinfo {year}
  {1984})}\BibitemShut {NoStop}%
\bibitem [{sem()}]{semi-Master}%
  \BibitemOpen
  \href@noop {} {}\bibinfo {note} {{A related conjecture has also been proposed
  in semiclassical chaotic
  systems~\cite{Berry77R,Bohigas84,Peres84E2,Feingold86,Grobe88,Srednicki94,Srednicki99,Muller04}.}}\BibitemShut
  {Stop}%
\bibitem [{\citenamefont {Karthik}\ \emph {et~al.}(2007)\citenamefont
  {Karthik}, \citenamefont {Sharma},\ and\ \citenamefont
  {Lakshminarayan}}]{Karthik07}%
  \BibitemOpen
  \bibfield  {author} {\bibinfo {author} {\bibfnamefont {J.}~\bibnamefont
  {Karthik}}, \bibinfo {author} {\bibfnamefont {A.}~\bibnamefont {Sharma}}, \
  and\ \bibinfo {author} {\bibfnamefont {A.}~\bibnamefont {Lakshminarayan}},\
  }\href {\doibase 10.1103/PhysRevA.75.022304} {\bibfield  {journal} {\bibinfo
  {journal} {Phys. Rev. A}\ }\textbf {\bibinfo {volume} {75}},\ \bibinfo
  {pages} {022304} (\bibinfo {year} {2007})}\BibitemShut {NoStop}%
\bibitem [{\citenamefont {Santos}\ and\ \citenamefont
  {Rigol}(2010{\natexlab{a}})}]{Santos10a}%
  \BibitemOpen
  \bibfield  {author} {\bibinfo {author} {\bibfnamefont {L.~F.}\ \bibnamefont
  {Santos}}\ and\ \bibinfo {author} {\bibfnamefont {M.}~\bibnamefont {Rigol}},\
  }\href {\doibase 10.1103/PhysRevE.81.036206} {\bibfield  {journal} {\bibinfo
  {journal} {Phys. Rev. E}\ }\textbf {\bibinfo {volume} {81}},\ \bibinfo
  {pages} {036206} (\bibinfo {year} {2010}{\natexlab{a}})}\BibitemShut
  {NoStop}%
\bibitem [{\citenamefont {Pal}\ and\ \citenamefont {Huse}(2010)}]{Pal10}%
  \BibitemOpen
  \bibfield  {author} {\bibinfo {author} {\bibfnamefont {A.}~\bibnamefont
  {Pal}}\ and\ \bibinfo {author} {\bibfnamefont {D.~A.}\ \bibnamefont {Huse}},\
  }\href {\doibase 10.1103/PhysRevB.82.174411} {\bibfield  {journal} {\bibinfo
  {journal} {Phys. Rev. B}\ }\textbf {\bibinfo {volume} {82}},\ \bibinfo
  {pages} {174411} (\bibinfo {year} {2010})}\BibitemShut {NoStop}%
\bibitem [{\citenamefont {Khatami}\ \emph {et~al.}(2013)\citenamefont
  {Khatami}, \citenamefont {Pupillo}, \citenamefont {Srednicki},\ and\
  \citenamefont {Rigol}}]{Khatami13}%
  \BibitemOpen
  \bibfield  {author} {\bibinfo {author} {\bibfnamefont {E.}~\bibnamefont
  {Khatami}}, \bibinfo {author} {\bibfnamefont {G.}~\bibnamefont {Pupillo}},
  \bibinfo {author} {\bibfnamefont {M.}~\bibnamefont {Srednicki}}, \ and\
  \bibinfo {author} {\bibfnamefont {M.}~\bibnamefont {Rigol}},\ }\href
  {\doibase 10.1103/PhysRevLett.111.050403} {\bibfield  {journal} {\bibinfo
  {journal} {Phys. Rev. Lett.}\ }\textbf {\bibinfo {volume} {111}},\ \bibinfo
  {pages} {050403} (\bibinfo {year} {2013})}\BibitemShut {NoStop}%
\bibitem [{\citenamefont {Beugeling}\ \emph {et~al.}(2014)\citenamefont
  {Beugeling}, \citenamefont {Moessner},\ and\ \citenamefont
  {Haque}}]{Beugeling14}%
  \BibitemOpen
  \bibfield  {author} {\bibinfo {author} {\bibfnamefont {W.}~\bibnamefont
  {Beugeling}}, \bibinfo {author} {\bibfnamefont {R.}~\bibnamefont {Moessner}},
  \ and\ \bibinfo {author} {\bibfnamefont {M.}~\bibnamefont {Haque}},\ }\href
  {\doibase 10.1103/PhysRevE.89.042112} {\bibfield  {journal} {\bibinfo
  {journal} {Phys. Rev. E}\ }\textbf {\bibinfo {volume} {89}},\ \bibinfo
  {pages} {042112} (\bibinfo {year} {2014})}\BibitemShut {NoStop}%
\bibitem [{\citenamefont {Kim}\ \emph {et~al.}(2014)\citenamefont {Kim},
  \citenamefont {Ikeda},\ and\ \citenamefont {Huse}}]{Kim14}%
  \BibitemOpen
  \bibfield  {author} {\bibinfo {author} {\bibfnamefont {H.}~\bibnamefont
  {Kim}}, \bibinfo {author} {\bibfnamefont {T.~N.}\ \bibnamefont {Ikeda}}, \
  and\ \bibinfo {author} {\bibfnamefont {D.~A.}\ \bibnamefont {Huse}},\ }\href
  {\doibase 10.1103/PhysRevE.90.052105} {\bibfield  {journal} {\bibinfo
  {journal} {Phys. Rev. E}\ }\textbf {\bibinfo {volume} {90}},\ \bibinfo
  {pages} {052105} (\bibinfo {year} {2014})}\BibitemShut {NoStop}%
\bibitem [{\citenamefont {Hamazaki}\ \emph {et~al.}(2016)\citenamefont
  {Hamazaki}, \citenamefont {Ikeda},\ and\ \citenamefont {Ueda}}]{Hamazaki16G}%
  \BibitemOpen
  \bibfield  {author} {\bibinfo {author} {\bibfnamefont {R.}~\bibnamefont
  {Hamazaki}}, \bibinfo {author} {\bibfnamefont {T.~N.}\ \bibnamefont {Ikeda}},
  \ and\ \bibinfo {author} {\bibfnamefont {M.}~\bibnamefont {Ueda}},\ }\href
  {\doibase 10.1103/PhysRevE.93.032116} {\bibfield  {journal} {\bibinfo
  {journal} {Phys. Rev. E}\ }\textbf {\bibinfo {volume} {93}},\ \bibinfo
  {pages} {032116} (\bibinfo {year} {2016})}\BibitemShut {NoStop}%
\bibitem [{\citenamefont {Beugeling}\ \emph {et~al.}(2015)\citenamefont
  {Beugeling}, \citenamefont {Moessner},\ and\ \citenamefont
  {Haque}}]{Beugeling15}%
  \BibitemOpen
  \bibfield  {author} {\bibinfo {author} {\bibfnamefont {W.}~\bibnamefont
  {Beugeling}}, \bibinfo {author} {\bibfnamefont {R.}~\bibnamefont {Moessner}},
  \ and\ \bibinfo {author} {\bibfnamefont {M.}~\bibnamefont {Haque}},\ }\href
  {\doibase 10.1103/PhysRevE.91.012144} {\bibfield  {journal} {\bibinfo
  {journal} {Phys. Rev. E}\ }\textbf {\bibinfo {volume} {91}},\ \bibinfo
  {pages} {012144} (\bibinfo {year} {2015})}\BibitemShut {NoStop}%
\bibitem [{\citenamefont {D'Alessio}\ \emph {et~al.}(2016)\citenamefont
  {D'Alessio}, \citenamefont {Kafri}, \citenamefont {Polkovnikov},\ and\
  \citenamefont {Rigol}}]{DAlessio16}%
  \BibitemOpen
  \bibfield  {author} {\bibinfo {author} {\bibfnamefont {L.}~\bibnamefont
  {D'Alessio}}, \bibinfo {author} {\bibfnamefont {Y.}~\bibnamefont {Kafri}},
  \bibinfo {author} {\bibfnamefont {A.}~\bibnamefont {Polkovnikov}}, \ and\
  \bibinfo {author} {\bibfnamefont {M.}~\bibnamefont {Rigol}},\ }\href
  {https://www.tandfonline.com/doi/abs/10.1080/00018732.2016.1198134}
  {\bibfield  {journal} {\bibinfo  {journal} {Advances in Physics}\ }\textbf
  {\bibinfo {volume} {65}},\ \bibinfo {pages} {239} (\bibinfo {year}
  {2016})}\BibitemShut {NoStop}%
\bibitem [{\citenamefont {Bera}\ and\ \citenamefont
  {Lakshminarayan}(2016)}]{Bera16}%
  \BibitemOpen
  \bibfield  {author} {\bibinfo {author} {\bibfnamefont {S.}~\bibnamefont
  {Bera}}\ and\ \bibinfo {author} {\bibfnamefont {A.}~\bibnamefont
  {Lakshminarayan}},\ }\href {\doibase 10.1103/PhysRevB.93.134204} {\bibfield
  {journal} {\bibinfo  {journal} {Phys. Rev. B}\ }\textbf {\bibinfo {volume}
  {93}},\ \bibinfo {pages} {134204} (\bibinfo {year} {2016})}\BibitemShut
  {NoStop}%
\bibitem [{\citenamefont {Luitz}\ and\ \citenamefont
  {Bar~Lev}(2016)}]{Luitz16}%
  \BibitemOpen
  \bibfield  {author} {\bibinfo {author} {\bibfnamefont {D.~J.}\ \bibnamefont
  {Luitz}}\ and\ \bibinfo {author} {\bibfnamefont {Y.}~\bibnamefont
  {Bar~Lev}},\ }\href {\doibase 10.1103/PhysRevLett.117.170404} {\bibfield
  {journal} {\bibinfo  {journal} {Phys. Rev. Lett.}\ }\textbf {\bibinfo
  {volume} {117}},\ \bibinfo {pages} {170404} (\bibinfo {year}
  {2016})}\BibitemShut {NoStop}%
\bibitem [{\citenamefont {Serbyn}\ and\ \citenamefont
  {Moore}(2016)}]{Serbyn16S}%
  \BibitemOpen
  \bibfield  {author} {\bibinfo {author} {\bibfnamefont {M.}~\bibnamefont
  {Serbyn}}\ and\ \bibinfo {author} {\bibfnamefont {J.~E.}\ \bibnamefont
  {Moore}},\ }\href {\doibase 10.1103/PhysRevB.93.041424} {\bibfield  {journal}
  {\bibinfo  {journal} {Phys. Rev. B}\ }\textbf {\bibinfo {volume} {93}},\
  \bibinfo {pages} {041424} (\bibinfo {year} {2016})}\BibitemShut {NoStop}%
\bibitem [{\citenamefont {Mondaini}\ \emph {et~al.}(2016)\citenamefont
  {Mondaini}, \citenamefont {Fratus}, \citenamefont {Srednicki},\ and\
  \citenamefont {Rigol}}]{Mondaini16}%
  \BibitemOpen
  \bibfield  {author} {\bibinfo {author} {\bibfnamefont {R.}~\bibnamefont
  {Mondaini}}, \bibinfo {author} {\bibfnamefont {K.~R.}\ \bibnamefont
  {Fratus}}, \bibinfo {author} {\bibfnamefont {M.}~\bibnamefont {Srednicki}}, \
  and\ \bibinfo {author} {\bibfnamefont {M.}~\bibnamefont {Rigol}},\ }\href
  {\doibase 10.1103/PhysRevE.93.032104} {\bibfield  {journal} {\bibinfo
  {journal} {Phys. Rev. E}\ }\textbf {\bibinfo {volume} {93}},\ \bibinfo
  {pages} {032104} (\bibinfo {year} {2016})}\BibitemShut {NoStop}%
\bibitem [{\citenamefont {Mondaini}\ and\ \citenamefont
  {Rigol}(2017)}]{Mondaini17}%
  \BibitemOpen
  \bibfield  {author} {\bibinfo {author} {\bibfnamefont {R.}~\bibnamefont
  {Mondaini}}\ and\ \bibinfo {author} {\bibfnamefont {M.}~\bibnamefont
  {Rigol}},\ }\href {\doibase 10.1103/PhysRevE.96.012157} {\bibfield  {journal}
  {\bibinfo  {journal} {Phys. Rev. E}\ }\textbf {\bibinfo {volume} {96}},\
  \bibinfo {pages} {012157} (\bibinfo {year} {2017})}\BibitemShut {NoStop}%
\bibitem [{\citenamefont {Pal}\ and\ \citenamefont
  {Lakshminarayan}(2018)}]{Pal18}%
  \BibitemOpen
  \bibfield  {author} {\bibinfo {author} {\bibfnamefont {R.}~\bibnamefont
  {Pal}}\ and\ \bibinfo {author} {\bibfnamefont {A.}~\bibnamefont
  {Lakshminarayan}},\ }\href {\doibase 10.1103/PhysRevB.98.174304} {\bibfield
  {journal} {\bibinfo  {journal} {Phys. Rev. B}\ }\textbf {\bibinfo {volume}
  {98}},\ \bibinfo {pages} {174304} (\bibinfo {year} {2018})}\BibitemShut
  {NoStop}%
\bibitem [{\citenamefont {Khaymovich}\ \emph {et~al.}(2018)\citenamefont
  {Khaymovich}, \citenamefont {Haque},\ and\ \citenamefont
  {McClarty}}]{Khaymovich18}%
  \BibitemOpen
  \bibfield  {author} {\bibinfo {author} {\bibfnamefont {I.~M.}\ \bibnamefont
  {Khaymovich}}, \bibinfo {author} {\bibfnamefont {M.}~\bibnamefont {Haque}}, \
  and\ \bibinfo {author} {\bibfnamefont {P.~A.}\ \bibnamefont {McClarty}},\
  }\href {https://arxiv.org/abs/1806.09631} {\bibfield  {journal} {\bibinfo
  {journal} {arXiv preprint arXiv:1806.09631}\ } (\bibinfo {year}
  {2018})}\BibitemShut {NoStop}%
\bibitem [{\citenamefont {Hamazaki}\ \emph
  {et~al.}(2018{\natexlab{a}})\citenamefont {Hamazaki}, \citenamefont
  {Kawabata},\ and\ \citenamefont {Ueda}}]{Hamazaki18N}%
  \BibitemOpen
  \bibfield  {author} {\bibinfo {author} {\bibfnamefont {R.}~\bibnamefont
  {Hamazaki}}, \bibinfo {author} {\bibfnamefont {K.}~\bibnamefont {Kawabata}},
  \ and\ \bibinfo {author} {\bibfnamefont {M.}~\bibnamefont {Ueda}},\ }\href
  {https://arxiv.org/abs/1811.11319} {\bibfield  {journal} {\bibinfo  {journal}
  {arXiv preprint arXiv:1811.11319}\ } (\bibinfo {year}
  {2018}{\natexlab{a}})}\BibitemShut {NoStop}%
\bibitem [{\citenamefont {Erd{\H{o}}s}\ and\ \citenamefont
  {Schr{\"o}der}(2014)}]{Erdos14}%
  \BibitemOpen
  \bibfield  {author} {\bibinfo {author} {\bibfnamefont {L.}~\bibnamefont
  {Erd{\H{o}}s}}\ and\ \bibinfo {author} {\bibfnamefont {D.}~\bibnamefont
  {Schr{\"o}der}},\ }\href
  {https://link.springer.com/article/10.1007/s11040-014-9164-3} {\bibfield
  {journal} {\bibinfo  {journal} {Mathematical Physics, Analysis and Geometry}\
  }\textbf {\bibinfo {volume} {17}},\ \bibinfo {pages} {441} (\bibinfo {year}
  {2014})}\BibitemShut {NoStop}%
\bibitem [{\citenamefont {Keating}\ \emph {et~al.}(2014)\citenamefont
  {Keating}, \citenamefont {Linden},\ and\ \citenamefont {Wells}}]{Keating14}%
  \BibitemOpen
  \bibfield  {author} {\bibinfo {author} {\bibfnamefont {J.}~\bibnamefont
  {Keating}}, \bibinfo {author} {\bibfnamefont {N.}~\bibnamefont {Linden}}, \
  and\ \bibinfo {author} {\bibfnamefont {H.}~\bibnamefont {Wells}},\ }\href
  {https://arxiv.org/abs/1403.1114} {\bibfield  {journal} {\bibinfo  {journal}
  {arXiv preprint arXiv:1403.1114}\ } (\bibinfo {year} {2014})}\BibitemShut
  {NoStop}%
\bibitem [{\citenamefont {Cunden}\ \emph {et~al.}(2017)\citenamefont {Cunden},
  \citenamefont {Maltsev},\ and\ \citenamefont {Mezzadri}}]{Cunden17}%
  \BibitemOpen
  \bibfield  {author} {\bibinfo {author} {\bibfnamefont {F.~D.}\ \bibnamefont
  {Cunden}}, \bibinfo {author} {\bibfnamefont {A.}~\bibnamefont {Maltsev}}, \
  and\ \bibinfo {author} {\bibfnamefont {F.}~\bibnamefont {Mezzadri}},\ }\href
  {https://aip.scitation.org/doi/abs/10.1063/1.4984942} {\bibfield  {journal}
  {\bibinfo  {journal} {Journal of Mathematical Physics}\ }\textbf {\bibinfo
  {volume} {58}},\ \bibinfo {pages} {061902} (\bibinfo {year}
  {2017})}\BibitemShut {NoStop}%
\bibitem [{\citenamefont {Kos}\ \emph {et~al.}(2018)\citenamefont {Kos},
  \citenamefont {Ljubotina},\ and\ \citenamefont {Prosen}}]{Kos17}%
  \BibitemOpen
  \bibfield  {author} {\bibinfo {author} {\bibfnamefont {P.}~\bibnamefont
  {Kos}}, \bibinfo {author} {\bibfnamefont {M.}~\bibnamefont {Ljubotina}}, \
  and\ \bibinfo {author} {\bibfnamefont {T.}~\bibnamefont {Prosen}},\ }\href
  {\doibase 10.1103/PhysRevX.8.021062} {\bibfield  {journal} {\bibinfo
  {journal} {Phys. Rev. X}\ }\textbf {\bibinfo {volume} {8}},\ \bibinfo {pages}
  {021062} (\bibinfo {year} {2018})}\BibitemShut {NoStop}%
\bibitem [{\citenamefont {Chan}\ \emph {et~al.}(2018)\citenamefont {Chan},
  \citenamefont {De~Luca},\ and\ \citenamefont {Chalker}}]{Chan18}%
  \BibitemOpen
  \bibfield  {author} {\bibinfo {author} {\bibfnamefont {A.}~\bibnamefont
  {Chan}}, \bibinfo {author} {\bibfnamefont {A.}~\bibnamefont {De~Luca}}, \
  and\ \bibinfo {author} {\bibfnamefont {J.~T.}\ \bibnamefont {Chalker}},\
  }\href {\doibase 10.1103/PhysRevLett.121.060601} {\bibfield  {journal}
  {\bibinfo  {journal} {Phys. Rev. Lett.}\ }\textbf {\bibinfo {volume} {121}},\
  \bibinfo {pages} {060601} (\bibinfo {year} {2018})}\BibitemShut {NoStop}%
\bibitem [{\citenamefont {Bertini}\ \emph {et~al.}(2018)\citenamefont
  {Bertini}, \citenamefont {Kos},\ and\ \citenamefont {Prosen}}]{Bertini18}%
  \BibitemOpen
  \bibfield  {author} {\bibinfo {author} {\bibfnamefont {B.}~\bibnamefont
  {Bertini}}, \bibinfo {author} {\bibfnamefont {P.}~\bibnamefont {Kos}}, \ and\
  \bibinfo {author} {\bibfnamefont {T.}~\bibnamefont {Prosen}},\ }\href
  {\doibase 10.1103/PhysRevLett.121.264101} {\bibfield  {journal} {\bibinfo
  {journal} {Phys. Rev. Lett.}\ }\textbf {\bibinfo {volume} {121}},\ \bibinfo
  {pages} {264101} (\bibinfo {year} {2018})}\BibitemShut {NoStop}%
\bibitem [{\citenamefont {Neumann}(1929)}]{Neumann29}%
  \BibitemOpen
  \bibfield  {author} {\bibinfo {author} {\bibfnamefont {J.~v.}\ \bibnamefont
  {Neumann}},\ }\href@noop {} {\bibfield  {journal} {\bibinfo  {journal}
  {Zeitschrift f{\"u}r Physik}\ }\textbf {\bibinfo {volume} {57}},\ \bibinfo
  {pages} {30} (\bibinfo {year} {1929})},\ \Eprint
  {http://arxiv.org/abs/English translation (by R. Tumulka), The European
  Physical Journal H 35, 201 (2010)} {English translation (by R. Tumulka), The
  European Physical Journal H 35, 201 (2010)} \BibitemShut {NoStop}%
\bibitem [{\citenamefont {Jensen}\ and\ \citenamefont
  {Shankar}(1985)}]{Jensen85}%
  \BibitemOpen
  \bibfield  {author} {\bibinfo {author} {\bibfnamefont {R.~V.}\ \bibnamefont
  {Jensen}}\ and\ \bibinfo {author} {\bibfnamefont {R.}~\bibnamefont
  {Shankar}},\ }\href {\doibase 10.1103/PhysRevLett.54.1879} {\bibfield
  {journal} {\bibinfo  {journal} {Phys. Rev. Lett.}\ }\textbf {\bibinfo
  {volume} {54}},\ \bibinfo {pages} {1879} (\bibinfo {year}
  {1985})}\BibitemShut {NoStop}%
\bibitem [{\citenamefont {Deutsch}(1991)}]{Deutsch91}%
  \BibitemOpen
  \bibfield  {author} {\bibinfo {author} {\bibfnamefont {J.~M.}\ \bibnamefont
  {Deutsch}},\ }\href {\doibase 10.1103/PhysRevA.43.2046} {\bibfield  {journal}
  {\bibinfo  {journal} {Phys. Rev. A}\ }\textbf {\bibinfo {volume} {43}},\
  \bibinfo {pages} {2046} (\bibinfo {year} {1991})}\BibitemShut {NoStop}%
\bibitem [{\citenamefont {Srednicki}(1994)}]{Srednicki94}%
  \BibitemOpen
  \bibfield  {author} {\bibinfo {author} {\bibfnamefont {M.}~\bibnamefont
  {Srednicki}},\ }\href {\doibase 10.1103/PhysRevE.50.888} {\bibfield
  {journal} {\bibinfo  {journal} {Phys. Rev. E}\ }\textbf {\bibinfo {volume}
  {50}},\ \bibinfo {pages} {888} (\bibinfo {year} {1994})}\BibitemShut
  {NoStop}%
\bibitem [{\citenamefont {Tasaki}(1998)}]{Tasaki98}%
  \BibitemOpen
  \bibfield  {author} {\bibinfo {author} {\bibfnamefont {H.}~\bibnamefont
  {Tasaki}},\ }\href {\doibase 10.1103/PhysRevLett.80.1373} {\bibfield
  {journal} {\bibinfo  {journal} {Phys. Rev. Lett.}\ }\textbf {\bibinfo
  {volume} {80}},\ \bibinfo {pages} {1373} (\bibinfo {year}
  {1998})}\BibitemShut {NoStop}%
\bibitem [{\citenamefont {Rigol}\ \emph {et~al.}(2008)\citenamefont {Rigol},
  \citenamefont {Dunjko},\ and\ \citenamefont {Olshanii}}]{Rigol08}%
  \BibitemOpen
  \bibfield  {author} {\bibinfo {author} {\bibfnamefont {M.}~\bibnamefont
  {Rigol}}, \bibinfo {author} {\bibfnamefont {V.}~\bibnamefont {Dunjko}}, \
  and\ \bibinfo {author} {\bibfnamefont {M.}~\bibnamefont {Olshanii}},\ }\href
  {https://www.nature.com/articles/nature06838} {\bibfield  {journal} {\bibinfo
   {journal} {Nature}\ }\textbf {\bibinfo {volume} {452}},\ \bibinfo {pages}
  {854} (\bibinfo {year} {2008})}\BibitemShut {NoStop}%
\bibitem [{\citenamefont {Santos}\ and\ \citenamefont
  {Rigol}(2010{\natexlab{b}})}]{Santos10b}%
  \BibitemOpen
  \bibfield  {author} {\bibinfo {author} {\bibfnamefont {L.~F.}\ \bibnamefont
  {Santos}}\ and\ \bibinfo {author} {\bibfnamefont {M.}~\bibnamefont {Rigol}},\
  }\href {\doibase 10.1103/PhysRevE.82.031130} {\bibfield  {journal} {\bibinfo
  {journal} {Phys. Rev. E}\ }\textbf {\bibinfo {volume} {82}},\ \bibinfo
  {pages} {031130} (\bibinfo {year} {2010}{\natexlab{b}})}\BibitemShut
  {NoStop}%
\bibitem [{\citenamefont {Ikeda}\ \emph {et~al.}(2011)\citenamefont {Ikeda},
  \citenamefont {Watanabe},\ and\ \citenamefont {Ueda}}]{Ikeda11}%
  \BibitemOpen
  \bibfield  {author} {\bibinfo {author} {\bibfnamefont {T.~N.}\ \bibnamefont
  {Ikeda}}, \bibinfo {author} {\bibfnamefont {Y.}~\bibnamefont {Watanabe}}, \
  and\ \bibinfo {author} {\bibfnamefont {M.}~\bibnamefont {Ueda}},\ }\href
  {\doibase 10.1103/PhysRevE.84.021130} {\bibfield  {journal} {\bibinfo
  {journal} {Phys. Rev. E}\ }\textbf {\bibinfo {volume} {84}},\ \bibinfo
  {pages} {021130} (\bibinfo {year} {2011})}\BibitemShut {NoStop}%
\bibitem [{\citenamefont {Khodja}\ \emph {et~al.}(2015)\citenamefont {Khodja},
  \citenamefont {Steinigeweg},\ and\ \citenamefont {Gemmer}}]{Khodja15}%
  \BibitemOpen
  \bibfield  {author} {\bibinfo {author} {\bibfnamefont {A.}~\bibnamefont
  {Khodja}}, \bibinfo {author} {\bibfnamefont {R.}~\bibnamefont {Steinigeweg}},
  \ and\ \bibinfo {author} {\bibfnamefont {J.}~\bibnamefont {Gemmer}},\ }\href
  {\doibase 10.1103/PhysRevE.91.012120} {\bibfield  {journal} {\bibinfo
  {journal} {Phys. Rev. E}\ }\textbf {\bibinfo {volume} {91}},\ \bibinfo
  {pages} {012120} (\bibinfo {year} {2015})}\BibitemShut {NoStop}%
\bibitem [{\citenamefont {Kaufman}\ \emph {et~al.}(2016)\citenamefont
  {Kaufman}, \citenamefont {Tai}, \citenamefont {Lukin}, \citenamefont
  {Rispoli}, \citenamefont {Schittko}, \citenamefont {Preiss},\ and\
  \citenamefont {Greiner}}]{Kaufman16}%
  \BibitemOpen
  \bibfield  {author} {\bibinfo {author} {\bibfnamefont {A.~M.}\ \bibnamefont
  {Kaufman}}, \bibinfo {author} {\bibfnamefont {M.~E.}\ \bibnamefont {Tai}},
  \bibinfo {author} {\bibfnamefont {A.}~\bibnamefont {Lukin}}, \bibinfo
  {author} {\bibfnamefont {M.}~\bibnamefont {Rispoli}}, \bibinfo {author}
  {\bibfnamefont {R.}~\bibnamefont {Schittko}}, \bibinfo {author}
  {\bibfnamefont {P.~M.}\ \bibnamefont {Preiss}}, \ and\ \bibinfo {author}
  {\bibfnamefont {M.}~\bibnamefont {Greiner}},\ }\href {\doibase
  10.1126/science.aaf6725} {\bibfield  {journal} {\bibinfo  {journal}
  {Science}\ }\textbf {\bibinfo {volume} {353}},\ \bibinfo {pages} {794}
  (\bibinfo {year} {2016})}\BibitemShut {NoStop}%
\bibitem [{\citenamefont {Clos}\ \emph {et~al.}(2016)\citenamefont {Clos},
  \citenamefont {Porras}, \citenamefont {Warring},\ and\ \citenamefont
  {Schaetz}}]{Clos16}%
  \BibitemOpen
  \bibfield  {author} {\bibinfo {author} {\bibfnamefont {G.}~\bibnamefont
  {Clos}}, \bibinfo {author} {\bibfnamefont {D.}~\bibnamefont {Porras}},
  \bibinfo {author} {\bibfnamefont {U.}~\bibnamefont {Warring}}, \ and\
  \bibinfo {author} {\bibfnamefont {T.}~\bibnamefont {Schaetz}},\ }\href
  {\doibase 10.1103/PhysRevLett.117.170401} {\bibfield  {journal} {\bibinfo
  {journal} {Phys. Rev. Lett.}\ }\textbf {\bibinfo {volume} {117}},\ \bibinfo
  {pages} {170401} (\bibinfo {year} {2016})}\BibitemShut {NoStop}%
\bibitem [{\citenamefont {Reimann}(2016)}]{Reimann16}%
  \BibitemOpen
  \bibfield  {author} {\bibinfo {author} {\bibfnamefont {P.}~\bibnamefont
  {Reimann}},\ }\href {https://www.nature.com/articles/ncomms10821} {\bibfield
  {journal} {\bibinfo  {journal} {Nature Communications}\ }\textbf {\bibinfo
  {volume} {7}} (\bibinfo {year} {2016})}\BibitemShut {NoStop}%
\bibitem [{\citenamefont {Mori}\ \emph {et~al.}(2018)\citenamefont {Mori},
  \citenamefont {Ikeda}, \citenamefont {Kaminishi},\ and\ \citenamefont
  {Ueda}}]{Mori18}%
  \BibitemOpen
  \bibfield  {author} {\bibinfo {author} {\bibfnamefont {T.}~\bibnamefont
  {Mori}}, \bibinfo {author} {\bibfnamefont {T.~N.}\ \bibnamefont {Ikeda}},
  \bibinfo {author} {\bibfnamefont {E.}~\bibnamefont {Kaminishi}}, \ and\
  \bibinfo {author} {\bibfnamefont {M.}~\bibnamefont {Ueda}},\ }\href
  {http://iopscience.iop.org/article/10.1088/1361-6455/aabcdf/meta} {\bibfield
  {journal} {\bibinfo  {journal} {Journal of Physics B: Atomic, Molecular and
  Optical Physics}\ }\textbf {\bibinfo {volume} {51}},\ \bibinfo {pages}
  {112001} (\bibinfo {year} {2018})}\BibitemShut {NoStop}%
\bibitem [{\citenamefont {Huang}\ \emph {et~al.}(2017)\citenamefont {Huang},
  \citenamefont {Brandao},\ and\ \citenamefont {Zhang}}]{Huang17}%
  \BibitemOpen
  \bibfield  {author} {\bibinfo {author} {\bibfnamefont {Y.}~\bibnamefont
  {Huang}}, \bibinfo {author} {\bibfnamefont {F.~G.}\ \bibnamefont {Brandao}},
  \ and\ \bibinfo {author} {\bibfnamefont {Y.-L.}\ \bibnamefont {Zhang}},\
  }\href {https://arxiv.org/abs/1705.07597} {\bibfield  {journal} {\bibinfo
  {journal} {arXiv preprint arXiv:1705.07597}\ } (\bibinfo {year}
  {2017})}\BibitemShut {NoStop}%
\bibitem [{\citenamefont {Schmitt}\ and\ \citenamefont
  {Kehrein}(2018)}]{Schmitt17}%
  \BibitemOpen
  \bibfield  {author} {\bibinfo {author} {\bibfnamefont {M.}~\bibnamefont
  {Schmitt}}\ and\ \bibinfo {author} {\bibfnamefont {S.}~\bibnamefont
  {Kehrein}},\ }\href {\doibase 10.1103/PhysRevB.98.180301} {\bibfield
  {journal} {\bibinfo  {journal} {Phys. Rev. B}\ }\textbf {\bibinfo {volume}
  {98}},\ \bibinfo {pages} {180301} (\bibinfo {year} {2018})}\BibitemShut
  {NoStop}%
\bibitem [{\citenamefont {Hamazaki}\ \emph
  {et~al.}(2018{\natexlab{b}})\citenamefont {Hamazaki}, \citenamefont
  {Fujimoto},\ and\ \citenamefont {Ueda}}]{Hamazaki18O}%
  \BibitemOpen
  \bibfield  {author} {\bibinfo {author} {\bibfnamefont {R.}~\bibnamefont
  {Hamazaki}}, \bibinfo {author} {\bibfnamefont {K.}~\bibnamefont {Fujimoto}},
  \ and\ \bibinfo {author} {\bibfnamefont {M.}~\bibnamefont {Ueda}},\ }\href
  {https://arxiv.org/abs/1807.02360} {\bibfield  {journal} {\bibinfo  {journal}
  {arXiv preprint arXiv:1807.02360}\ } (\bibinfo {year}
  {2018}{\natexlab{b}})}\BibitemShut {NoStop}%
\bibitem [{\citenamefont {Dyson}(1962)}]{Dyson62T}%
  \BibitemOpen
  \bibfield  {author} {\bibinfo {author} {\bibfnamefont {F.~J.}\ \bibnamefont
  {Dyson}},\ }\href {https://aip.scitation.org/doi/abs/10.1063/1.1703863}
  {\bibfield  {journal} {\bibinfo  {journal} {Journal of Mathematical Physics}\
  }\textbf {\bibinfo {volume} {3}},\ \bibinfo {pages} {1199} (\bibinfo {year}
  {1962})}\BibitemShut {NoStop}%
\bibitem [{\citenamefont {Ba\~nuls}\ \emph {et~al.}(2011)\citenamefont
  {Ba\~nuls}, \citenamefont {Cirac},\ and\ \citenamefont
  {Hastings}}]{Banuls11}%
  \BibitemOpen
  \bibfield  {author} {\bibinfo {author} {\bibfnamefont {M.~C.}\ \bibnamefont
  {Ba\~nuls}}, \bibinfo {author} {\bibfnamefont {J.~I.}\ \bibnamefont {Cirac}},
  \ and\ \bibinfo {author} {\bibfnamefont {M.~B.}\ \bibnamefont {Hastings}},\
  }\href {\doibase 10.1103/PhysRevLett.106.050405} {\bibfield  {journal}
  {\bibinfo  {journal} {Phys. Rev. Lett.}\ }\textbf {\bibinfo {volume} {106}},\
  \bibinfo {pages} {050405} (\bibinfo {year} {2011})}\BibitemShut {NoStop}%
\bibitem [{\citenamefont {Shenker}\ and\ \citenamefont
  {Stanford}(2014)}]{Shenker14}%
  \BibitemOpen
  \bibfield  {author} {\bibinfo {author} {\bibfnamefont {S.~H.}\ \bibnamefont
  {Shenker}}\ and\ \bibinfo {author} {\bibfnamefont {D.}~\bibnamefont
  {Stanford}},\ }\href
  {https://link.springer.com/article/10.1007/JHEP03(2014)067} {\bibfield
  {journal} {\bibinfo  {journal} {Journal of High Energy Physics}\ }\textbf
  {\bibinfo {volume} {2014}},\ \bibinfo {pages} {67} (\bibinfo {year}
  {2014})}\BibitemShut {NoStop}%
\bibitem [{\citenamefont {Zhang}\ \emph {et~al.}(2015)\citenamefont {Zhang},
  \citenamefont {Kim},\ and\ \citenamefont {Huse}}]{Zhang15}%
  \BibitemOpen
  \bibfield  {author} {\bibinfo {author} {\bibfnamefont {L.}~\bibnamefont
  {Zhang}}, \bibinfo {author} {\bibfnamefont {H.}~\bibnamefont {Kim}}, \ and\
  \bibinfo {author} {\bibfnamefont {D.~A.}\ \bibnamefont {Huse}},\ }\href
  {\doibase 10.1103/PhysRevE.91.062128} {\bibfield  {journal} {\bibinfo
  {journal} {Phys. Rev. E}\ }\textbf {\bibinfo {volume} {91}},\ \bibinfo
  {pages} {062128} (\bibinfo {year} {2015})}\BibitemShut {NoStop}%
\bibitem [{\citenamefont {Kim}\ \emph {et~al.}(2015)\citenamefont {Kim},
  \citenamefont {Ba\~nuls}, \citenamefont {Cirac}, \citenamefont {Hastings},\
  and\ \citenamefont {Huse}}]{Kim15}%
  \BibitemOpen
  \bibfield  {author} {\bibinfo {author} {\bibfnamefont {H.}~\bibnamefont
  {Kim}}, \bibinfo {author} {\bibfnamefont {M.~C.}\ \bibnamefont {Ba\~nuls}},
  \bibinfo {author} {\bibfnamefont {J.~I.}\ \bibnamefont {Cirac}}, \bibinfo
  {author} {\bibfnamefont {M.~B.}\ \bibnamefont {Hastings}}, \ and\ \bibinfo
  {author} {\bibfnamefont {D.~A.}\ \bibnamefont {Huse}},\ }\href {\doibase
  10.1103/PhysRevE.92.012128} {\bibfield  {journal} {\bibinfo  {journal} {Phys.
  Rev. E}\ }\textbf {\bibinfo {volume} {92}},\ \bibinfo {pages} {012128}
  (\bibinfo {year} {2015})}\BibitemShut {NoStop}%
\bibitem [{\citenamefont {Hosur}\ \emph {et~al.}(2016)\citenamefont {Hosur},
  \citenamefont {Qi}, \citenamefont {Roberts},\ and\ \citenamefont
  {Yoshida}}]{Hosur16}%
  \BibitemOpen
  \bibfield  {author} {\bibinfo {author} {\bibfnamefont {P.}~\bibnamefont
  {Hosur}}, \bibinfo {author} {\bibfnamefont {X.-L.}\ \bibnamefont {Qi}},
  \bibinfo {author} {\bibfnamefont {D.~A.}\ \bibnamefont {Roberts}}, \ and\
  \bibinfo {author} {\bibfnamefont {B.}~\bibnamefont {Yoshida}},\ }\href
  {https://link.springer.com/article/10.1007/JHEP02(2016)004} {\bibfield
  {journal} {\bibinfo  {journal} {Journal of High Energy Physics}\ }\textbf
  {\bibinfo {volume} {2016}},\ \bibinfo {pages} {4} (\bibinfo {year}
  {2016})}\BibitemShut {NoStop}%
\bibitem [{\citenamefont {Smith}\ \emph {et~al.}(2016)\citenamefont {Smith},
  \citenamefont {Lee}, \citenamefont {Richerme}, \citenamefont {Neyenhuis},
  \citenamefont {Hess}, \citenamefont {Hauke}, \citenamefont {Heyl},
  \citenamefont {Huse},\ and\ \citenamefont {Monroe}}]{Smith16}%
  \BibitemOpen
  \bibfield  {author} {\bibinfo {author} {\bibfnamefont {J.}~\bibnamefont
  {Smith}}, \bibinfo {author} {\bibfnamefont {A.}~\bibnamefont {Lee}}, \bibinfo
  {author} {\bibfnamefont {P.}~\bibnamefont {Richerme}}, \bibinfo {author}
  {\bibfnamefont {B.}~\bibnamefont {Neyenhuis}}, \bibinfo {author}
  {\bibfnamefont {P.~W.}\ \bibnamefont {Hess}}, \bibinfo {author}
  {\bibfnamefont {P.}~\bibnamefont {Hauke}}, \bibinfo {author} {\bibfnamefont
  {M.}~\bibnamefont {Heyl}}, \bibinfo {author} {\bibfnamefont {D.~A.}\
  \bibnamefont {Huse}}, \ and\ \bibinfo {author} {\bibfnamefont
  {C.}~\bibnamefont {Monroe}},\ }\href {\doibase 10.1038/nphys3783} {\bibfield
  {journal} {\bibinfo  {journal} {Nature Physics}\ }\textbf {\bibinfo {volume}
  {12}},\ \bibinfo {pages} {907} (\bibinfo {year} {2016})}\BibitemShut
  {NoStop}%
\bibitem [{add()}]{added-Master}%
  \BibitemOpen
  \href@noop {} {}\bibinfo {note} {{This work is based on RH's master
  thesis~\cite{HamazakiM}, which was submitted to the University of Tokyo in
  January 2017. Some results related to this master thesis were published
  afterwards~\cite{Mondaini17,Khaymovich18,Hamazaki18A}, as cited throughout
  this paper.}}\BibitemShut {Stop}%
\bibitem [{\citenamefont {Dzyaloshinsky}(1958)}]{Dzyaloshinsky58}%
  \BibitemOpen
  \bibfield  {author} {\bibinfo {author} {\bibfnamefont {I.}~\bibnamefont
  {Dzyaloshinsky}},\ }\href
  {https://www.sciencedirect.com/science/article/pii/0022369758900763}
  {\bibfield  {journal} {\bibinfo  {journal} {Journal of Physics and Chemistry
  of Solids}\ }\textbf {\bibinfo {volume} {4}},\ \bibinfo {pages} {241}
  (\bibinfo {year} {1958})}\BibitemShut {NoStop}%
\bibitem [{\citenamefont {Moriya}(1960)}]{Moriya60}%
  \BibitemOpen
  \bibfield  {author} {\bibinfo {author} {\bibfnamefont {T.}~\bibnamefont
  {Moriya}},\ }\href {\doibase 10.1103/PhysRev.120.91} {\bibfield  {journal}
  {\bibinfo  {journal} {Phys. Rev.}\ }\textbf {\bibinfo {volume} {120}},\
  \bibinfo {pages} {91} (\bibinfo {year} {1960})}\BibitemShut {NoStop}%
\bibitem [{\citenamefont {Haake}(2010)}]{Haake}%
  \BibitemOpen
  \bibfield  {author} {\bibinfo {author} {\bibfnamefont {F.}~\bibnamefont
  {Haake}},\ }\href {https://www.springer.com/us/book/9783642054273} {\emph
  {\bibinfo {title} {Quantum signatures of chaos}}},\ Vol.~\bibinfo {volume}
  {54}\ (\bibinfo  {publisher} {Springer Science \& Business Media},\ \bibinfo
  {year} {2010})\BibitemShut {NoStop}%
\bibitem [{\citenamefont {Atas}\ \emph
  {et~al.}(2013{\natexlab{a}})\citenamefont {Atas}, \citenamefont {Bogomolny},
  \citenamefont {Giraud},\ and\ \citenamefont {Roux}}]{Atas13}%
  \BibitemOpen
  \bibfield  {author} {\bibinfo {author} {\bibfnamefont {Y.~Y.}\ \bibnamefont
  {Atas}}, \bibinfo {author} {\bibfnamefont {E.}~\bibnamefont {Bogomolny}},
  \bibinfo {author} {\bibfnamefont {O.}~\bibnamefont {Giraud}}, \ and\ \bibinfo
  {author} {\bibfnamefont {G.}~\bibnamefont {Roux}},\ }\href {\doibase
  10.1103/PhysRevLett.110.084101} {\bibfield  {journal} {\bibinfo  {journal}
  {Phys. Rev. Lett.}\ }\textbf {\bibinfo {volume} {110}},\ \bibinfo {pages}
  {084101} (\bibinfo {year} {2013}{\natexlab{a}})}\BibitemShut {NoStop}%
\bibitem [{deg()}]{degen-Master}%
  \BibitemOpen
  \href@noop {} {}\bibinfo {note} {{For example,
  $\ket{\uparrow\downarrow\uparrow\cdots\downarrow\uparrow\uparrow}$ and
  $\ket{\downarrow\uparrow\downarrow\cdots\uparrow\downarrow\downarrow}$ are
  energy eigenstates of $\hat{H}_\mr{I}$ having the same energy. Here,
  $\ket{\uparrow}$ and $\ket{\downarrow}$ are eigenstates of $\hat{\sigma}^z$
  with eigenvalues $+1$ or $-1$, respectively.}}\BibitemShut {Stop}%
\bibitem [{\citenamefont {Dymarsky}\ and\ \citenamefont
  {Liu}(2019)}]{Dymarsky19}%
  \BibitemOpen
  \bibfield  {author} {\bibinfo {author} {\bibfnamefont {A.}~\bibnamefont
  {Dymarsky}}\ and\ \bibinfo {author} {\bibfnamefont {H.}~\bibnamefont {Liu}},\
  }\href {\doibase 10.1103/PhysRevE.99.010102} {\bibfield  {journal} {\bibinfo
  {journal} {Phys. Rev. E}\ }\textbf {\bibinfo {volume} {99}},\ \bibinfo
  {pages} {010102} (\bibinfo {year} {2019})}\BibitemShut {NoStop}%
\bibitem [{\citenamefont {Srednicki}(1999)}]{Srednicki99}%
  \BibitemOpen
  \bibfield  {author} {\bibinfo {author} {\bibfnamefont {M.}~\bibnamefont
  {Srednicki}},\ }\href {\doibase 10.1088/0305-4470/32/7/007} {\bibfield
  {journal} {\bibinfo  {journal} {Journal of Physics A: Mathematical and
  General}\ }\textbf {\bibinfo {volume} {32}},\ \bibinfo {pages} {1163}
  (\bibinfo {year} {1999})}\BibitemShut {NoStop}%
\bibitem [{\citenamefont {Hamazaki}\ and\ \citenamefont
  {Ueda}(2018)}]{Hamazaki18A}%
  \BibitemOpen
  \bibfield  {author} {\bibinfo {author} {\bibfnamefont {R.}~\bibnamefont
  {Hamazaki}}\ and\ \bibinfo {author} {\bibfnamefont {M.}~\bibnamefont
  {Ueda}},\ }\href {\doibase 10.1103/PhysRevLett.120.080603} {\bibfield
  {journal} {\bibinfo  {journal} {Phys. Rev. Lett.}\ }\textbf {\bibinfo
  {volume} {120}},\ \bibinfo {pages} {080603} (\bibinfo {year}
  {2018})}\BibitemShut {NoStop}%
\bibitem [{\citenamefont {Steinigeweg}\ \emph {et~al.}(2013)\citenamefont
  {Steinigeweg}, \citenamefont {Herbrych},\ and\ \citenamefont
  {Prelov\ifmmode~\check{s}\else \v{s}\fi{}ek}}]{Steinigeweg13}%
  \BibitemOpen
  \bibfield  {author} {\bibinfo {author} {\bibfnamefont {R.}~\bibnamefont
  {Steinigeweg}}, \bibinfo {author} {\bibfnamefont {J.}~\bibnamefont
  {Herbrych}}, \ and\ \bibinfo {author} {\bibfnamefont {P.}~\bibnamefont
  {Prelov\ifmmode~\check{s}\else \v{s}\fi{}ek}},\ }\href {\doibase
  10.1103/PhysRevE.87.012118} {\bibfield  {journal} {\bibinfo  {journal} {Phys.
  Rev. E}\ }\textbf {\bibinfo {volume} {87}},\ \bibinfo {pages} {012118}
  (\bibinfo {year} {2013})}\BibitemShut {NoStop}%
\bibitem [{\citenamefont {Ikeda}(2015)}]{IkedaD}%
  \BibitemOpen
  \bibfield  {author} {\bibinfo {author} {\bibfnamefont {T.}~\bibnamefont
  {Ikeda}},\ }\href {http://hdl.handle.net/2261/60163} {\enquote {\bibinfo
  {title} {Theoretical study on the foundation of statistical mechanics in
  isolated quantum systems},}\ } (\bibinfo {year} {2015}),\ \bibinfo {note}
  {{Ph.D. thesis}}\BibitemShut {NoStop}%
\bibitem [{\citenamefont {Dymarsky}(2018)}]{Dymarsky18}%
  \BibitemOpen
  \bibfield  {author} {\bibinfo {author} {\bibfnamefont {A.}~\bibnamefont
  {Dymarsky}},\ }\href {https://arxiv.org/abs/1804.08626} {\bibfield  {journal}
  {\bibinfo  {journal} {arXiv preprint arXiv:1804.08626}\ } (\bibinfo {year}
  {2018})}\BibitemShut {NoStop}%
\bibitem [{bre()}]{breit-Master}%
  \BibitemOpen
  \href@noop {} {}\bibinfo {note} {{The Breit-Wigner width is the energy width
  that characterizes the strength of the symmetry-breaking
  perturbation~\cite{Guhr98}.}}\BibitemShut {Stop}%
\bibitem [{exc()}]{exc-Master}%
  \BibitemOpen
  \href@noop {} {}\bibinfo {note} {{It is possible that $\Delta E_\mr{Univ}$
  exceeds the Thouless energy and the Breit-Wigner width for many-body
  observables. Indeed, most random observables in the entire operator space
  (including non-local many-body observables) exhibit the universal ratio
  without any limitation on the energy scale~\cite{Hamazaki18A}.}}\BibitemShut
  {Stop}%
\bibitem [{\citenamefont {Islam}\ \emph {et~al.}(2015)\citenamefont {Islam},
  \citenamefont {Ma}, \citenamefont {Preiss}, \citenamefont {Tai},
  \citenamefont {Lukin}, \citenamefont {Rispoli},\ and\ \citenamefont
  {Greiner}}]{Islam15}%
  \BibitemOpen
  \bibfield  {author} {\bibinfo {author} {\bibfnamefont {R.}~\bibnamefont
  {Islam}}, \bibinfo {author} {\bibfnamefont {R.}~\bibnamefont {Ma}}, \bibinfo
  {author} {\bibfnamefont {P.~M.}\ \bibnamefont {Preiss}}, \bibinfo {author}
  {\bibfnamefont {M.~E.}\ \bibnamefont {Tai}}, \bibinfo {author} {\bibfnamefont
  {A.}~\bibnamefont {Lukin}}, \bibinfo {author} {\bibfnamefont
  {M.}~\bibnamefont {Rispoli}}, \ and\ \bibinfo {author} {\bibfnamefont
  {M.}~\bibnamefont {Greiner}},\ }\href
  {https://www.nature.com/articles/nature15750} {\bibfield  {journal} {\bibinfo
   {journal} {Nature}\ }\textbf {\bibinfo {volume} {528}},\ \bibinfo {pages}
  {77} (\bibinfo {year} {2015})}\BibitemShut {NoStop}%
\bibitem [{\citenamefont {Feingold}\ and\ \citenamefont
  {Peres}(1986)}]{Feingold86}%
  \BibitemOpen
  \bibfield  {author} {\bibinfo {author} {\bibfnamefont {M.}~\bibnamefont
  {Feingold}}\ and\ \bibinfo {author} {\bibfnamefont {A.}~\bibnamefont
  {Peres}},\ }\href {\doibase 10.1103/PhysRevA.34.591} {\bibfield  {journal}
  {\bibinfo  {journal} {Phys. Rev. A}\ }\textbf {\bibinfo {volume} {34}},\
  \bibinfo {pages} {591} (\bibinfo {year} {1986})}\BibitemShut {NoStop}%
\bibitem [{\citenamefont {Reimann}(2008)}]{Reimann08}%
  \BibitemOpen
  \bibfield  {author} {\bibinfo {author} {\bibfnamefont {P.}~\bibnamefont
  {Reimann}},\ }\href {\doibase 10.1103/PhysRevLett.101.190403} {\bibfield
  {journal} {\bibinfo  {journal} {Phys. Rev. Lett.}\ }\textbf {\bibinfo
  {volume} {101}},\ \bibinfo {pages} {190403} (\bibinfo {year}
  {2008})}\BibitemShut {NoStop}%
\bibitem [{\citenamefont {Guhr}\ \emph {et~al.}(1998)\citenamefont {Guhr},
  \citenamefont {M{\"u}ller-Groeling},\ and\ \citenamefont
  {Weidenm{\"u}ller}}]{Guhr98}%
  \BibitemOpen
  \bibfield  {author} {\bibinfo {author} {\bibfnamefont {T.}~\bibnamefont
  {Guhr}}, \bibinfo {author} {\bibfnamefont {A.}~\bibnamefont
  {M{\"u}ller-Groeling}}, \ and\ \bibinfo {author} {\bibfnamefont {H.~A.}\
  \bibnamefont {Weidenm{\"u}ller}},\ }\href
  {https://www.sciencedirect.com/science/article/pii/S0370157397000884}
  {\bibfield  {journal} {\bibinfo  {journal} {Physics Reports}\ }\textbf
  {\bibinfo {volume} {299}},\ \bibinfo {pages} {189} (\bibinfo {year}
  {1998})}\BibitemShut {NoStop}%
\bibitem [{\citenamefont {Atas}\ \emph
  {et~al.}(2013{\natexlab{b}})\citenamefont {Atas}, \citenamefont {Bogomolny},
  \citenamefont {Giraud}, \citenamefont {Vivo},\ and\ \citenamefont
  {Vivo}}]{Atas13J}%
  \BibitemOpen
  \bibfield  {author} {\bibinfo {author} {\bibfnamefont {Y.}~\bibnamefont
  {Atas}}, \bibinfo {author} {\bibfnamefont {E.}~\bibnamefont {Bogomolny}},
  \bibinfo {author} {\bibfnamefont {O.}~\bibnamefont {Giraud}}, \bibinfo
  {author} {\bibfnamefont {P.}~\bibnamefont {Vivo}}, \ and\ \bibinfo {author}
  {\bibfnamefont {E.}~\bibnamefont {Vivo}},\ }\href@noop {} {\bibfield
  {journal} {\bibinfo  {journal} {Journal of Physics A: Mathematical and
  Theoretical}\ }\textbf {\bibinfo {volume} {46}},\ \bibinfo {pages} {355204}
  (\bibinfo {year} {2013}{\natexlab{b}})}\BibitemShut {NoStop}%
\bibitem [{\citenamefont {Hamazaki}(2017)}]{HamazakiM}%
  \BibitemOpen
  \bibfield  {author} {\bibinfo {author} {\bibfnamefont {R.}~\bibnamefont
  {Hamazaki}},\ }\href@noop {} {\enquote {\bibinfo {title} {Theoretical study
  on thermalization in isolated quantum systems},}\ } (\bibinfo {year}
  {2017}),\ \bibinfo {note} {{Master's thesis, arXiv:1901.01481}}\BibitemShut
  {NoStop}%
\bibitem [{\citenamefont {Berry}(1977)}]{Berry77R}%
  \BibitemOpen
  \bibfield  {author} {\bibinfo {author} {\bibfnamefont {M.~V.}\ \bibnamefont
  {Berry}},\ }\href
  {http://iopscience.iop.org/article/10.1088/0305-4470/10/12/016/meta}
  {\bibfield  {journal} {\bibinfo  {journal} {Journal of Physics A:
  Mathematical and General}\ }\textbf {\bibinfo {volume} {10}},\ \bibinfo
  {pages} {2083} (\bibinfo {year} {1977})}\BibitemShut {NoStop}%
\bibitem [{\citenamefont {Bohigas}\ \emph {et~al.}(1984)\citenamefont
  {Bohigas}, \citenamefont {Giannoni},\ and\ \citenamefont
  {Schmit}}]{Bohigas84}%
  \BibitemOpen
  \bibfield  {author} {\bibinfo {author} {\bibfnamefont {O.}~\bibnamefont
  {Bohigas}}, \bibinfo {author} {\bibfnamefont {M.~J.}\ \bibnamefont
  {Giannoni}}, \ and\ \bibinfo {author} {\bibfnamefont {C.}~\bibnamefont
  {Schmit}},\ }\href {\doibase 10.1103/PhysRevLett.52.1} {\bibfield  {journal}
  {\bibinfo  {journal} {Phys. Rev. Lett.}\ }\textbf {\bibinfo {volume} {52}},\
  \bibinfo {pages} {1} (\bibinfo {year} {1984})}\BibitemShut {NoStop}%
\bibitem [{\citenamefont {Feingold}\ \emph {et~al.}(1984)\citenamefont
  {Feingold}, \citenamefont {Moiseyev},\ and\ \citenamefont
  {Peres}}]{Peres84E2}%
  \BibitemOpen
  \bibfield  {author} {\bibinfo {author} {\bibfnamefont {M.}~\bibnamefont
  {Feingold}}, \bibinfo {author} {\bibfnamefont {N.}~\bibnamefont {Moiseyev}},
  \ and\ \bibinfo {author} {\bibfnamefont {A.}~\bibnamefont {Peres}},\ }\href
  {\doibase 10.1103/PhysRevA.30.509} {\bibfield  {journal} {\bibinfo  {journal}
  {Phys. Rev. A}\ }\textbf {\bibinfo {volume} {30}},\ \bibinfo {pages} {509}
  (\bibinfo {year} {1984})}\BibitemShut {NoStop}%
\bibitem [{\citenamefont {Grobe}\ \emph {et~al.}(1988)\citenamefont {Grobe},
  \citenamefont {Haake},\ and\ \citenamefont {Sommers}}]{Grobe88}%
  \BibitemOpen
  \bibfield  {author} {\bibinfo {author} {\bibfnamefont {R.}~\bibnamefont
  {Grobe}}, \bibinfo {author} {\bibfnamefont {F.}~\bibnamefont {Haake}}, \ and\
  \bibinfo {author} {\bibfnamefont {H.-J.}\ \bibnamefont {Sommers}},\ }\href
  {\doibase 10.1103/PhysRevLett.61.1899} {\bibfield  {journal} {\bibinfo
  {journal} {Phys. Rev. Lett.}\ }\textbf {\bibinfo {volume} {61}},\ \bibinfo
  {pages} {1899} (\bibinfo {year} {1988})}\BibitemShut {NoStop}%
\bibitem [{\citenamefont {M\"uller}\ \emph {et~al.}(2004)\citenamefont
  {M\"uller}, \citenamefont {Heusler}, \citenamefont {Braun}, \citenamefont
  {Haake},\ and\ \citenamefont {Altland}}]{Muller04}%
  \BibitemOpen
  \bibfield  {author} {\bibinfo {author} {\bibfnamefont {S.}~\bibnamefont
  {M\"uller}}, \bibinfo {author} {\bibfnamefont {S.}~\bibnamefont {Heusler}},
  \bibinfo {author} {\bibfnamefont {P.}~\bibnamefont {Braun}}, \bibinfo
  {author} {\bibfnamefont {F.}~\bibnamefont {Haake}}, \ and\ \bibinfo {author}
  {\bibfnamefont {A.}~\bibnamefont {Altland}},\ }\href {\doibase
  10.1103/PhysRevLett.93.014103} {\bibfield  {journal} {\bibinfo  {journal}
  {Phys. Rev. Lett.}\ }\textbf {\bibinfo {volume} {93}},\ \bibinfo {pages}
  {014103} (\bibinfo {year} {2004})}\BibitemShut {NoStop}%
\end{thebibliography}%

%\end{widetext}

\end{document}